%% file: hpp-paper.tex
\documentclass[generic]{imsart}

\RequirePackage{amsthm,amsmath,amsfonts,amssymb}
\RequirePackage[authoryear]{natbib}
\RequirePackage[colorlinks,citecolor=blue,urlcolor=blue]{hyperref}
\RequirePackage{graphicx}

\usepackage[a4paper,margin=1.25in]{geometry}

\startlocaldefs
\theoremstyle{plain}

\theoremstyle{remark}

\usepackage{mathtools}
\usepackage[capitalize,nameinlink]{cleveref}
\usepackage{subfig,subfloat}
\usepackage{algorithm}
\usepackage{algorithmicx}
\usepackage{algpseudocode}
\usepackage{listings}
\usepackage{color}
\usepackage{bbm}
\usepackage{tabularx,caption,float}
\usepackage{enumitem}

\input{defs.tex}

\graphicspath{{figures/}}

\endlocaldefs

\begin{document}

\begin{frontmatter}
\title{Modelling financial volume curves with hierarchical Poisson processes}
\runtitle{Hierarchical Poisson processes}

\begin{aug}
\author[A]{\fnms{Creighton}~\snm{Heaukulani}\ead[label=e1]{c.k.heaukulani@gmail.com}},
\author[B]{\fnms{Abhinav}~\snm{Pandey}\ead[label=e2]{abhinav.pandey@connect.ust.hk}}
\and
\author[B]{\fnms{Lancelot F.}~\snm{James}\ead[label=e3]{lancelot@ust.hk}}
\address[A]{Singapore\printead[presep={,\ }]{e1}}

\address[B]{Department of ISOM,
Hong Kong University of Science and Technology\printead[presep={,\ }]{e2,e3}}
\end{aug}

\begin{abstract}
Modeling the trading volume curves of financial instruments throughout the day is of key interest in financial trading applications. Predictions of these so-called volume profiles guide trade execution strategies, for example, a common strategy is to trade a desired quantity across many orders in line with the expected volume curve throughout the day so as not to impact the price of the instrument. 
The volume curves (for each day) are naturally grouped by stock and can be further gathered into higher-level groupings, such as by industry. 
In order to model such admixtures of volume curves, we introduce a hierarchical Poisson process model for the intensity functions of admixtures of inhomogenous Poisson processes, which represent the trading times of the stock throughout the day. The model is based on the hierarchical Dirichlet process, and an efficient Markov Chain Monte Carlo (MCMC) algorithm is derived following the slice sampling framework for Bayesian nonparametric mixture models.
We demonstrate the method on datasets of different stocks from the Trade and Quote repository maintained by Wharton Research Data Services, including the most liquid stock on the NASDAQ stock exchange, Apple, demonstrating the scalability of the approach.
\end{abstract}


\begin{keyword}
\kwd{Bayesian nonparametrics}
\kwd{machine learning}
\kwd{quantitative finance}
\kwd{intensity function}
\kwd{time series}
\end{keyword}

\end{frontmatter}

\section{Introduction and data}
\label{sec:intro}

In this article, we provide a generative model for multiple observations of an event-based time series. We will take as motivation the particular application of modeling the \emph{trading volume curves} of financial instruments, i.e., the volume of the instrument that is traded throughout each day over multiple days. For a particular instrument (such as a stock, bond, currency, or exchange traded fund), a trading volume curve takes the form of a jump process on a bounded interval. The interval represents the trading hours of the day, the curve jumps (in continuous time) whenever the stock trades, and the jump size is given by the amount of the stock that is traded. (The data will be described in depth and visualized in \cref{sec:data}.)
The classic approach to flexibly model a jump process is with an \emph{(inhomogenous) Poisson process} \citep{kingman1993}.
Indeed, we opt to model a single day's trading volume curve with the model for \emph{marked} Poisson processes described by \citet{taddy2012mixture}, whose intensity functions are constructed from Dirichlet process mixtures of beta density functions \citep{kottas2006dirichlet,kottas2007bayesian}.

In our application, the trading volume curve is recorded every day, and our dataset is therefore comprised of multiple time series recording the jump processes for each stock. In this work, we couple the Poisson processes for the different days (each characterized by their intensity functions, which are derived from Dirichlet process mixture models), through the \emph{hierarchical Dirichlet process} construction \citep{tehhdp}. We call the resulting model for the collection of (coupled) Poisson processes the \emph{hierarchical Poisson process}. We will also see that this hierarchy can be easily extended to multiple layers (by extending the hierarchical Dirichlet process to multiple layers), which naturally couples the models across different instruments/stocks, and may more generally capture broader groupings of the financial instruments, for example, into company sectors.

We implement a Markov Chain Monte Carlo (MCMC) algorithm to perform posterior inference in the model. 
We apply appropriate extensions of the \emph{auxiliary slice sampling methods} for Dirichlet process mixture models, described by \citet{walker2007sampling,kalli2011slice}, to the hierarchical Dirichlet process construction. 
Our algorithms rely on a representation for the \emph{Chinese restaurant franchise} due to \citet{tehhdp}, and provide novel extensions to the slice samplers for the hierarchical Dirichlet process studied, for example, by \citet{amini2019exact}.

Alternative Bayesian models for flexible, inhomogenous Poisson process intensity functions include the classic \emph{log Gaussian Cox process} \citep{moller1998log}, which models the intensity function as the log of a Gaussian process. There are other models that construct the intensity function from Gaussian processes, for example, \citet{adams2009tractable} maps the Gaussian process to positive values with a scaled sigmoid, and \citet{lloyd2015variational} squares the Gaussian process. And it should be noted that there are several existing Bayesian nonparametric approaches to the related task of modeling hazard curves and cumulative intensity functions \citep{ishwaran2004comp}.

Beyond finance, such \emph{event datasets}, i.e., time series of random, continuous-valued event times (often with corresponding measurements), are common. For example, geologists record the times throughout the day of geyser emissions in U.S. national parks, in order to model geothermal activity. 
In many of these applications, the event time series are recorded repeatedly, for example, the emissions from the geysers can be recorded every day over the course of a year. 
\citet{taddy2012mixture} show that their techniques may generalize to arbitrary applications where the data is appropriately modeled by a marked Poisson process on a bounded interval, and likewise our methods may naturally generalize to any application where there are collections of such time series, and where those collections may be additionally grouped into broader categories.

The remainder of the article is organized as follows. In \cref{sec:data}, we describe and visualize the dataset of trading volume curves in detail.
\cref{sec:model} describes our modeling methodology. In \cref{sec:timespp,sec:markedpp}, we model a single day's trading volume curve with the Poisson process framework, and in \cref{sec:hpp} we define the hierarchical Poisson process construction to couple across the models for different days and to capture broader groupings of the data.
In \cref{sec:inference}, we detail the MCMC procedure for posterior inference, and in \cref{sec:experiments} we provide some empirical demonstrations.

\subsection{Cumulative volume curves in financial trading}
\label{sec:data}

In a stock market, shares of stocks are traded in fixed quantities, called \emph{lots}. 
For any particular stock, traders may place a \emph{bid} (specifying a number of lots and the price they are willing to pay for those lots) when they want to buy the stock and an \emph{ask} (specifying a number of lots and the price they are willing to accept for those lots) when they want to sell the stock.
When a bid (quantity and price) matches an ask (quantity and price), a trade is executed by the stock exchange (the NASDAQ or NYSE, for example). 
The ways in which modern stock markets execute orders is rather more elaborate than this simple description, however, we will only focus on datasets of these standard trades during regular market hours, which form the bulk of the data collected and studied on intraday trading. This is reasonable since most actors in the financial industry model idiosyncratic stock market characteristics separately (such as trades executed off-hours, or during opening and closing bulk auctions). This will also allow us to highlight how the hierarchical Poisson process model that we introduce later can be generally applied to similar datasets beyond finance.
\begin{figure}[ht!]
\centering
\subfloat[AAPL - cumulative traded lots]{
	\includegraphics[scale=0.45]{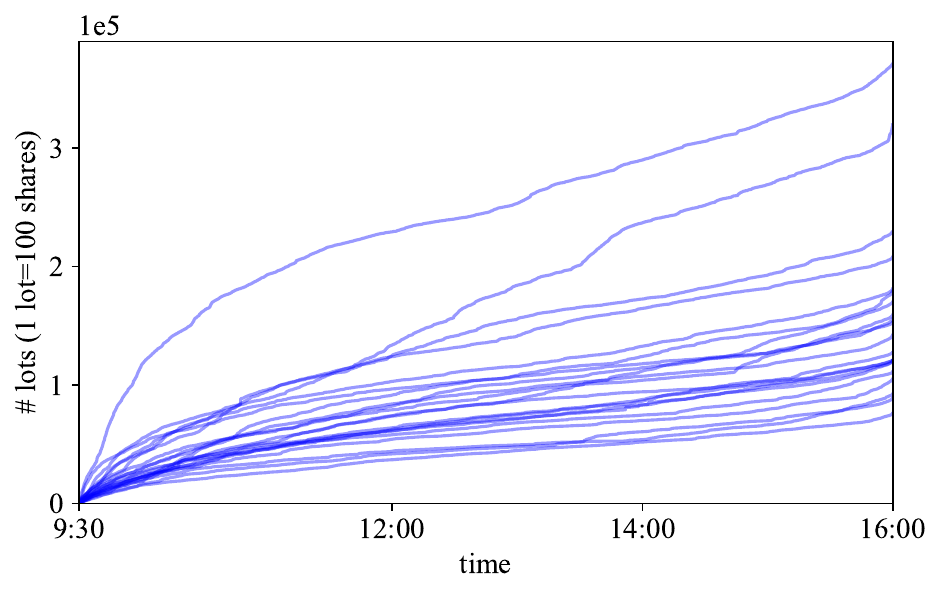}
	\label{fig:aapl}
}
\subfloat[AAPL - cumulative number of trades]{
	\includegraphics[scale=0.45]{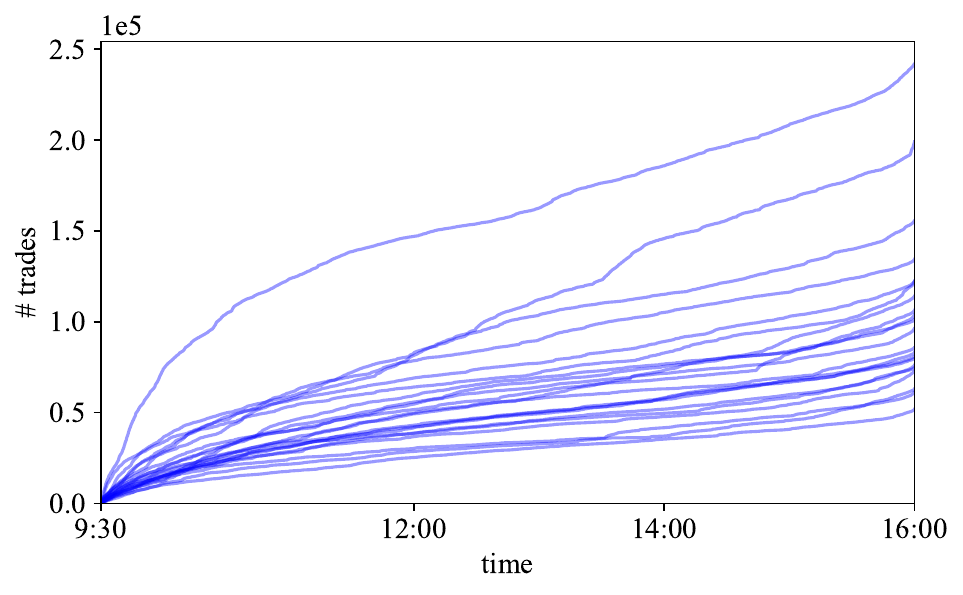}
	\label{fig:aapl_trades}
}\\
\subfloat[IRBT - cumulative traded lots]{
	\includegraphics[scale=0.45]{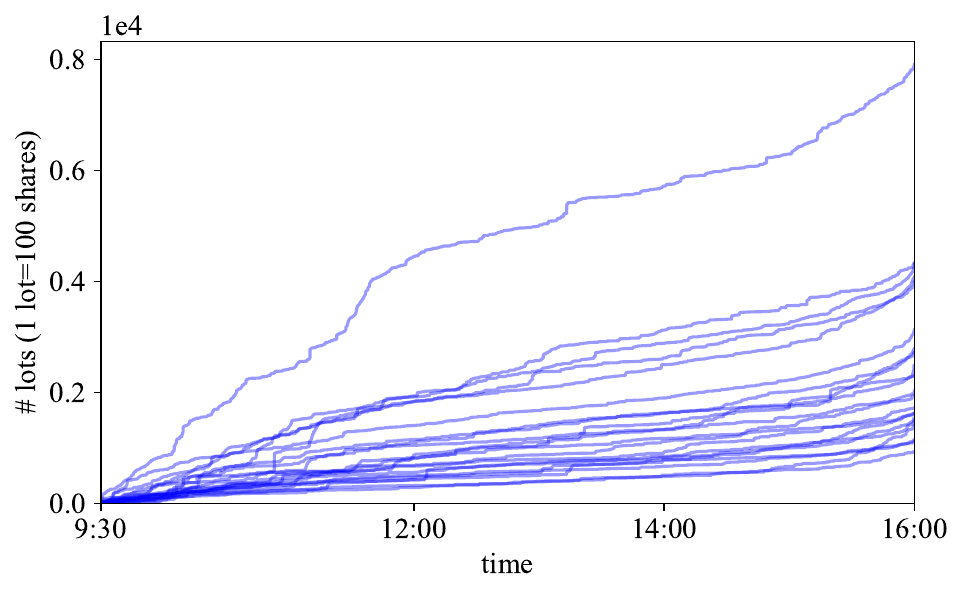}
	\label{fig:irbt}
}
\subfloat[IRBT - cumulative number of trades]{
	\includegraphics[scale=0.45]{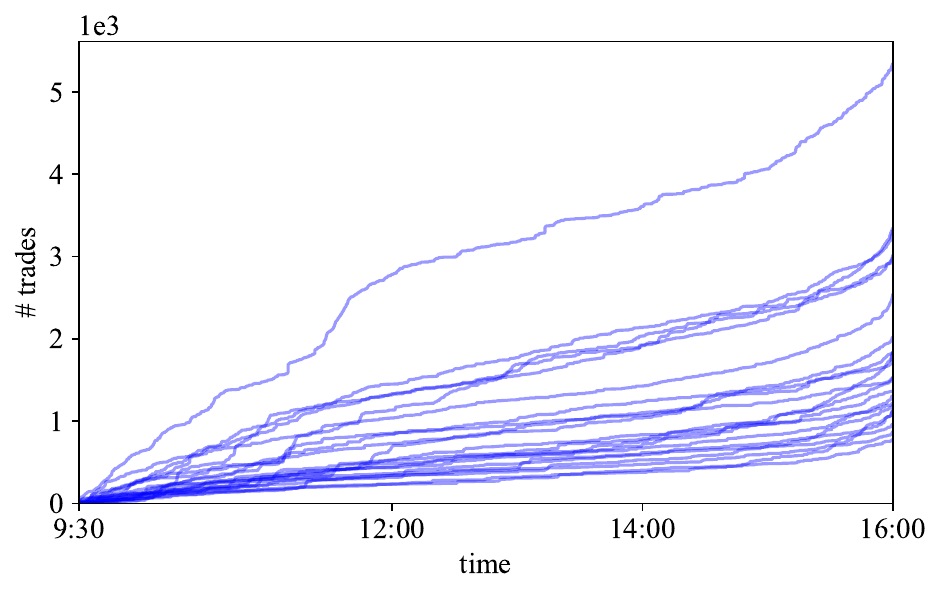}
	\label{fig:irbt_trades}
}\\
\subfloat[PZZA - cumulative traded lots]{
	\includegraphics[scale=0.45]{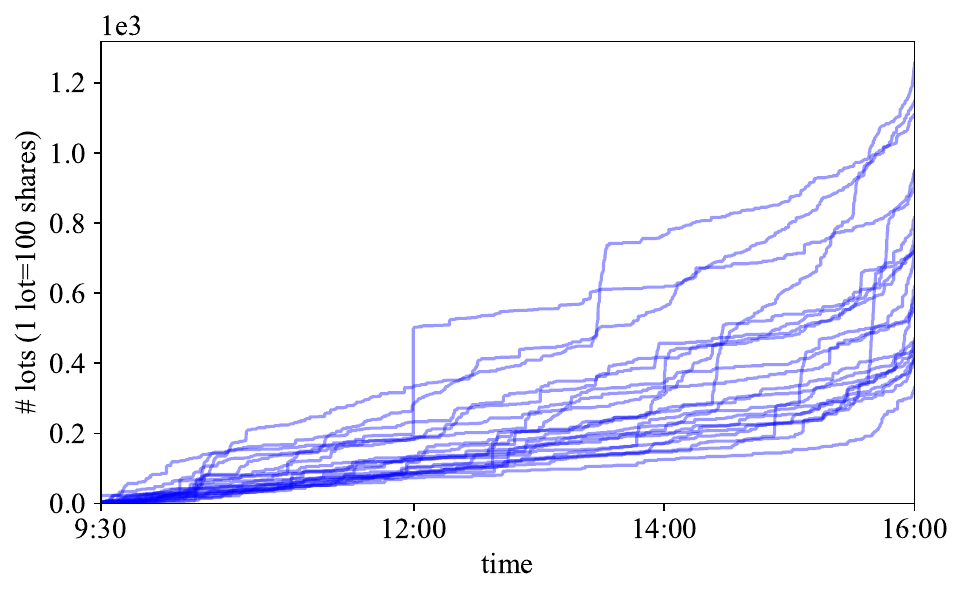}
	\label{fig:pzza}
}
\subfloat[PZZA - cumulative number of trades]{
	\includegraphics[scale=0.45]{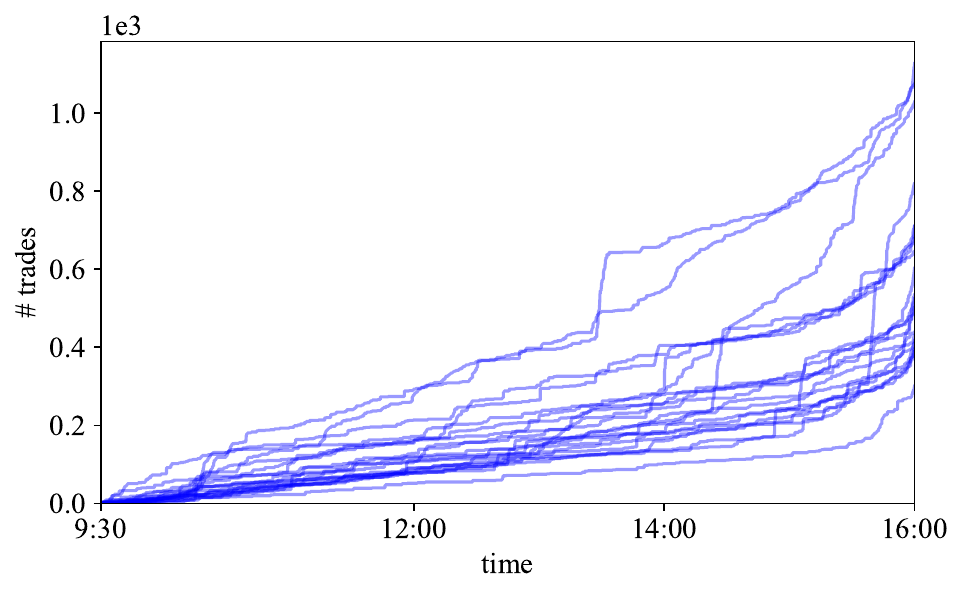}
	\label{fig:pzza_trades}
}
\caption{Trading volume curves for stocks in Apple (AAPL), iRobot (IRBT), and Papa John's Pizza (PZZA), which are well-known examples of large-, medium-, and small-cap stocks, respectively. Each curve represents a different day (there are 21 days in each dataset). {\bf Left column}: cumulative trading volume (in number of lots traded), which is a step function that jumps by the size of the trade. {\bf Right column}: the cumulative number of trades, which is a step function that jumps by one.}
\label{fig:trade-data}
\end{figure}

The datasets we will study are shown in \cref{fig:trade-data}; these plots display intraday cumulative trading activity, for a collection of days. (We provide details on how to recreate these datasets in \cref{sec:datanotes}.) 
In particular, \cref{fig:aapl} displays the cumulative volume (in lots) of Apple stock (NASDAQ symbol AAPL) traded during the market hours (9:30 -- 16:00 EST for the NASDAQ stock exchange) of each day in January 2013. Each curve represents a different trading day in the month (so there are 21 curves). Note that each curve is a step function that jumps every time a trade of Apple stock is executed, and the amount the curve increases is the number of lots traded. 
In what follows, we will first consider Poisson process models for the trading times alone, and we will subsequently mark the jumps of the Poisson process with the trade sizes.
To this end, it is helpful to additionally visualize the cumulative number of trades, displayed in \cref{fig:aapl_trades} for the AAPL dataset. Note that this step function simply jumps by a value of one at each trade time.

Apple, currently the worlds most valuable company, is an example of a \emph{large-capitalization} (a.k.a. \emph{large-cap}) company, which is a classification based on market capitalization and correspondingly describes the \emph{liquidity} of the stock, that is, how frequently the stock trades throughout the day. 
It will also be instructive to analyze less liquid financial instruments. 
In \cref{fig:irbt,fig:irbt_trades}, we display the trading volume curves for iRobot (NASDAQ symbol: IRBT), a well-known medium-cap stock for a company specializing in robotic vacuum cleaners, and in \cref{fig:pzza,fig:pzza_trades} we display the trading volume curves for Papa John's Pizza (NASDAQ symbol: PZZA), a well-known small-cap stock for the pizza restaurant franchise.
We immediately see some visual differences between the volume curves for the three stocks. The trading curves for AAPL, the most liquid stock, appear ``smoother'', reflecting the higher frequency of trading. The ``resolution'' of the curve degrades as the stock becomes less liquid, with IRBT and particularly PZZA having more noticeable abrupt jumps in the curve. The differences between the cumulative lots traded and the cumulative number of trades are also more noticeable as the stock becomes less liquid. These characteristics all highlight that investors fundamentally differ their trading strategy according to liquidity.

\subsection{Dataset details}
\label{sec:datanotes}

\begin{figure}[th!]
\centering
\subfloat[AAPL - smoothed histogram of trade times]{
	\includegraphics[scale=0.45]{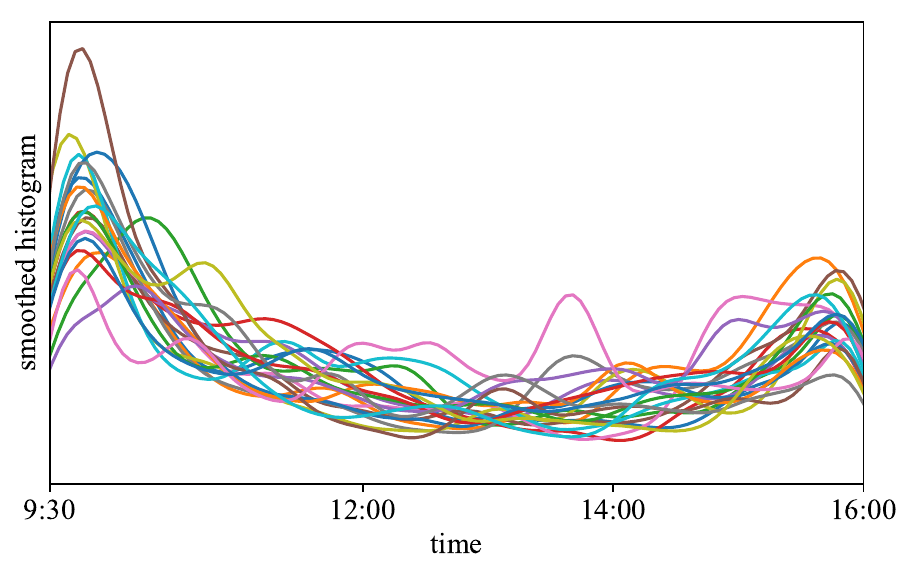}
	\label{fig:aapl_kde}
}
\subfloat[AAPL - boxplots of the \# of lots traded]{
	\includegraphics[scale=0.45]{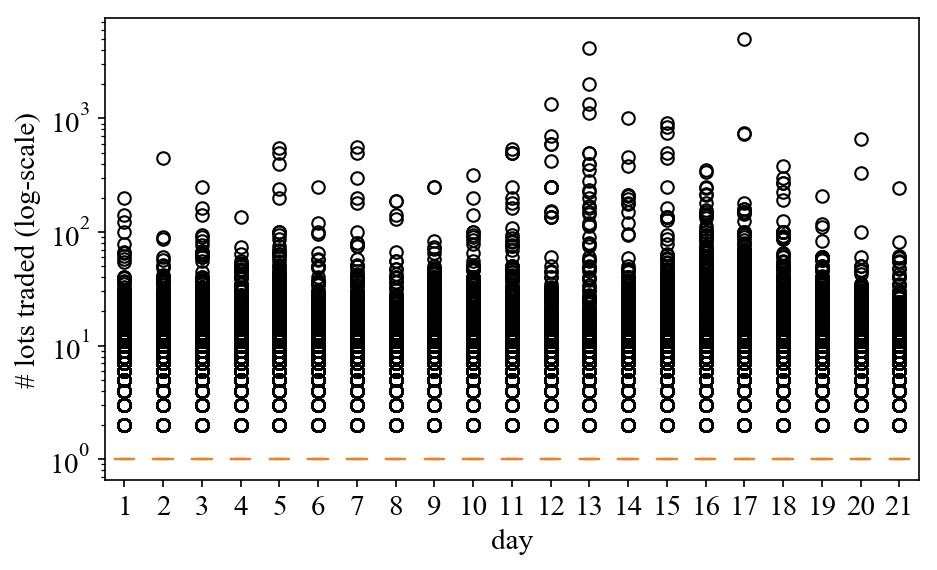}
	\label{fig:aapl_size}
}\\
\subfloat[IRBT - smoothed histogram of trade times]{
	\includegraphics[scale=0.45]{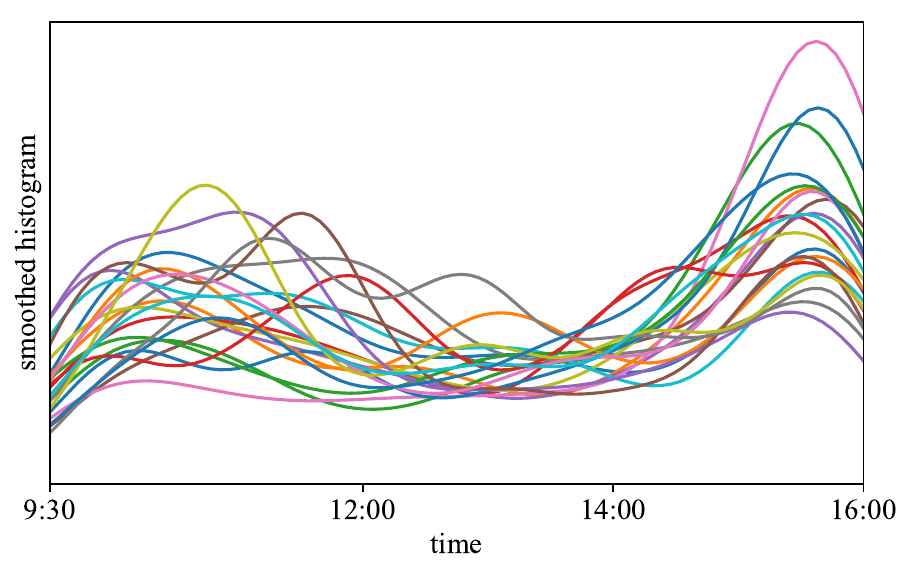}
	\label{fig:irbt_kde}
}
\subfloat[IRBT - boxplots of the \# of lots traded]{
	\includegraphics[scale=0.45]{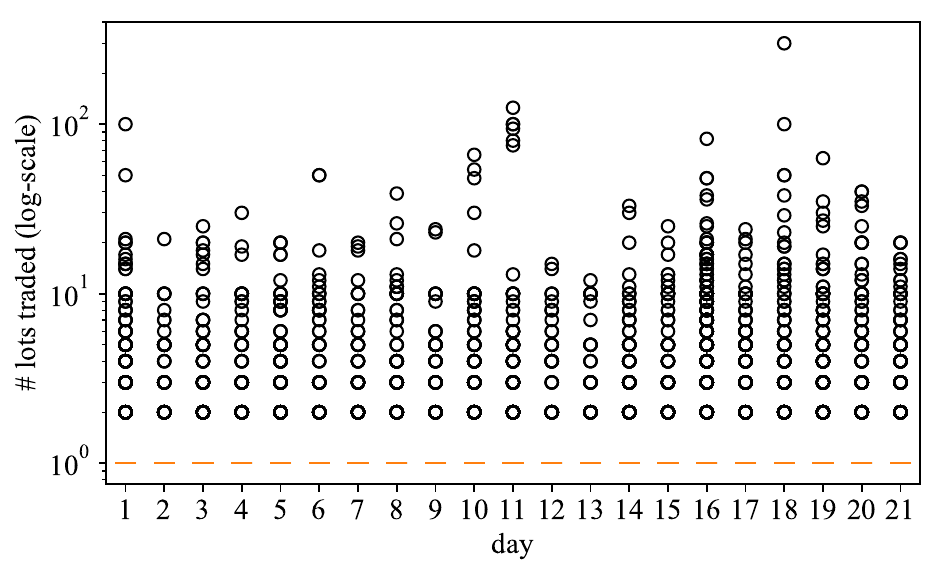}
	\label{fig:irbt_size}
}\\
\subfloat[PZZA - smoothed histogram of trade times]{
	\includegraphics[scale=0.45]{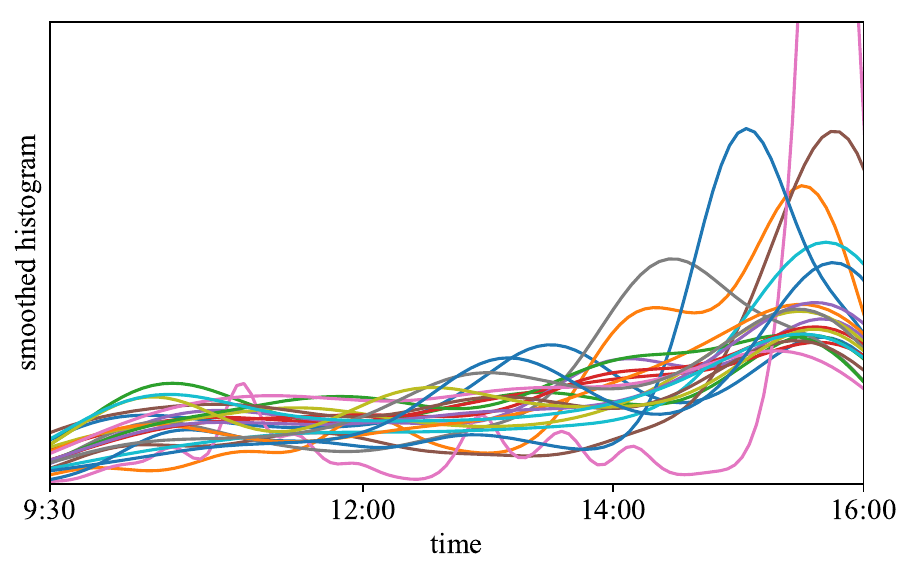}
	\label{fig:pzza_kde}
}
\subfloat[PZZA - boxplots of the \# of lots traded]{
	\includegraphics[scale=0.45]{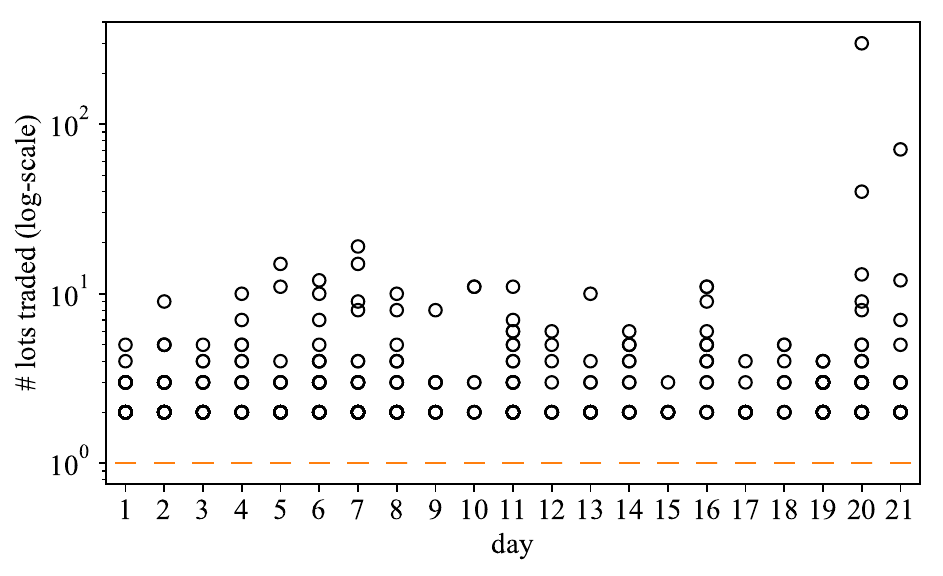}
	\label{fig:pzza_size}
}
\caption{{\bf Left column}: Smoothed histograms (truncated kernel density estimates) of the time of the trade. Each curve represents a different day. {\bf Right column}: Boxplots of the number of lots traded (i.e., the size of the trade), displayed on a log-scale. Each boxplot represents a different day. Note that almost all of the mass (the entire ``box'') is collapsed onto a single lot.}
\label{fig:data-details}
\end{figure}

\begin{table}[h!]
\centering
\begin{tabular}{| l | l | l | l | l | l |}
\hline
Dataset & Total trades & Trades / day & Min. trades & Max. trades & \% single lot \\
\hline \hline
{\bf AAPL} & 2,228,087 & 106,099 $\pm$ 45,313 & 242,086 & 51,826 & 81.46 \% \\
{\bf IRBT} & 41,074 & 1,956 $\pm$ 1,091 & 5,345 & 778 & 89.07 \% \\
{\bf PZZA} & 12,659 & 603 $\pm$ 233 & 1,127 & 303 & 93.66 \% \\
\hline
\end{tabular}
\caption{Statistics of the datasets. In order: the total number of trades, the (approximate) number of trades per day (mean $\pm$ one standard deviation), the number of trades on the least active day, the number of trades on the most active day, and the percentage of the trade sizes (across all days) that are a single lot.}
\label{tab:datastats}
\end{table}

Statistics of the datasets are shown in \cref{tab:datastats}, including the total number of trades (over the 21 days), the mean and standard deviation of the number of trades per day, and the number of trades on the least active and most active days.
Note that the medium-cap stock (IRBT) is about 3.2x as liquid as the small-cap stock (PZZA), and the large-cap stock (AAPL) is about 54.2x as liquid as IRBT and about 176.0x as liquid as PZZA.

To get an idea of the structure we aim to model, consider \cref{fig:aapl_kde}, which displays ``smoothed histograms'' of the trade times for each day in the AAPL dataset, derived from (truncated) kernel density estimates. As before, each curve represents a different day (the colors of which we have alternated to aid visualization), and we may interpret these curves as describing the relative likelihood of the time of a trade. Intuitively, this is what the \emph{intensity function} of an inhomogeneous Poisson process models, and in what follows, we essentially want to ``smooth'' over these different intensity functions by sharing structure among them. Similarly, the smoothed histograms for the trade times in the IRBT and PZZA datasets are displayed in \cref{fig:irbt_kde} and \cref{fig:pzza_kde}, respectively. Note that AAPL has relatively higher activity during the morning, IRBT is a bit more balanced between morning and afternoon, with slightly higher mass on the latter, and PZZA places most of its mass toward the end of the day. 

The distributions of trade sizes are slightly more difficult to visualize: they have a sharp peak at one lot (most trades are of a single lot) and have a very heavy tail. 
The most useful display is a boxplot (of the number of traded lots) displayed on a log-scale for each day, as shown in \cref{fig:aapl_size}, \cref{fig:irbt_size}, and \cref{fig:pzza_size} for the AAPL, IRBT, and PZZA datasets, respectively.
Note that by far most of the mass (the entire ``box'', in fact) is collapsed onto a single lot, and the thickness of the tail depends on liquidity. Note that we provide the fraction of trades that are of only a single lot in the last column of \cref{tab:datastats}.
This emphasizes that our primary goal should be to model the intensity function for the time of the trade correctly, since this inherently models all single lot trades, covering 80--90 \% of the data.

Finally, we detail how to obtain the datasets. The raw data was obtained from the Trade and Quote repository maintained by Wharton Research Data Services. The consolidated trade datasets were extracted (for the month of January 2013), and pre- and post-market hour trades were excluded (simply discard trades outside of 9:30--16:00 EST). \emph{Odd lot} trades were removed from the dataset, which are easily identified as those executions that are not multiples of 100 shares (the \emph{round lot} size for the NASDAQ market).
Unfortunately, the freely available data has a stored resolution of only up to one second. This is unrealistic of the data that is actually used by actors in the financial markets (purchased from a number of vendors), and so we simulate the true trade times by uniformly sampling the continuous-valued time within a one second window (i.e., $\pm$ 0.5 sec.) around the measurements in the raw dataset. At the microsecond level, there were no collisions in any of the datasets we study.

\section{Modeling methodology}
\label{sec:model}

\subsection{Bayesian nonparametric intensity functions on a bounded interval}
\label{sec:timespp}

Let $D$ denote the number of days (i.e., $D=21$ for the three datasets we study), and let $\PosInts \defas \{1, 2, \dots \}$ denote the positive integers. For every day $d \le D$, we represent the cumulative number of lots traded (as displayed in \cref{fig:aapl,fig:irbt,fig:pzza}) by a $\PosInts$-valued process $X_d$ indexed by the unit interval $(0,1)$. In particular, we map the market hours of the trading day to the unit interval, where zero denotes the start of the trading day (09:30 EST for NASDAQ), and one denotes the end of the trading day (16:00 EST for NASDAQ). 
For every $d\le D$, let $N_d$ denote the number of trades on day $d$, then we may write
\[
X_d ( \tau ) = \sum_{i=1}^{N_d} q_{d,i} \Indicator{ t_{d,i} \le \tau }
	,
	\qquad \tau \in (0,1)
	,
	\label{eq:eventdata}
\]
where $t_{d,i} \in (0,1)$ and $q_{d,i} \in \PosInts$ denote the time and quantity (in lots), respectively, of trade $i\le N_d$ on day $d\le D$. 
Note that the processes recording only the number of jumps (as displayed in \cref{fig:aapl_trades,fig:irbt_trades,fig:pzza_trades}), may be represented by
\[
\Ical_d(\tau) = \sum_{i=1}^{N_d} \Indicator{ t_{d,i} \le \tau }
	,
	\qquad \tau \in (0,1)
	,
	\label{eq:countprocess}
\]
which are \emph{counting processes} on $(0,1)$. In this article, we will define Poisson point processes to model $\Ical_d(\tau)$ and mark them to obtain the model for $X_d(\tau)$.

Fix $d \le D$.
Let $\lambda_d(t)$ be a nonnegative, locally integrable function on $(0,1)$ and model the counting process $\Ical_d(t)$ given by \cref{eq:countprocess} as a Poisson point process on $(0,1)$ with intensity function $\lambda_d(t)$. 
Then the number of jumps $N_d$ of the process $\Ical_d(t)$ on $(0,1)$ is Poisson distributed with mean $\Lambda_d \defas \int_0^1 \lambda_d(t) \dee t$, and, conditioned on $N_d$, the jump locations (i.e., the trade times) $t_d \defas (t_{d,1}, \dotsc, t_{d,N_d})$ are \iid\ with density function $f_d(t) \defas \lambda_d(t) / \Lambda_d$.
More formally,
\[
p ( t_d, N_d \given \Lambda_d ) \propto \Lambda_d^{N_d} \exp ( - \Lambda_d ) \prod_{i=1}^{N_d} f_d ( t_{d,i} )
	.
	\label{eq:countlikel}
\]
\citet{kottas2006dirichlet} noted that the factorization of the likelihood in \cref{eq:countlikel} implies that we may model $f_d(t)$ and $\Lambda_d$ independently. We follow their specification for these two objects: let $\Lambda_d \dist \gammadist(a_\Lambda, b_\Lambda)$ for some $a_\Lambda, b_\Lambda>0$, and let $f_d(t)$ be a nonparametric mixture of beta density functions. 
In particular, this mixture model has an infinite number of components, and the mixing weights are given by a \emph{Dirichlet process}.

Recall that a Dirichlet process $G$ on a Borel space $(\Theta, \Acal)$ with concentration parameter $\conc > 0$ and base measure $H$, denoted $G \dist \DPLAW(\conc, H)$, may be written as
\[
G = \sum_{j=1}^\infty \nu_j \delta_{\theta_j}
	,
\]
for a collection of random elements $\nu_1, \nu_2, \dots$ in $(0,1]$ satisfying $\sum_{j=1}^\infty = 1$ almost surely (\as), and a collection of random elements $\theta_1, \theta_2, \dots$ in $\Theta$ that are distributed \iid\ according to $H$.
The weights $\nu \defas (\nu_1, \nu_2, \dots)$ may be constructed via their \emph{stick-breaking} representation \citep{sethuraman1994constructive}, on which we will rely for inference:
\[
\nu_j' \dist \betadist(1, \conc_d ) 
	,
	\qquad
	\nu_j = \nu_j' \prod_{i=1}^{j-1} ( 1 - \nu_i' ) 
	,
	\qquad
	j = 1, 2, \dots
	.
	\label{eq:localsticks}
\]

Then we construct
\[
f_d (t; G_d) \defas \int \betadist( t ; \alpha, \beta ) \dee G_d ( \alpha, \beta )
	,
	\qquad
	G_d \dist \DPLAW ( \conc_d, H )
	,
	\label{eq:dpdensity}
\]
for some $\conc_d >0$ and a non-atomic measure $H$ that can be interpreted as a prior distribution on $\alpha, \beta > 0$. 
Note that we have expanded our notation for $f_d$ to emphasize dependence on the Dirichlet process $G_d$. 
We may then write the (random) intensity function of the Poisson process $\Ical_d(t)$ as
\[
\lambda_d (t; G_d) = \Lambda_d \int \betadist( t ; \alpha, \beta ) \dee G_d ( \alpha, \beta )
	.
	\label{eq:countintensity}
\]
In particular, note that $G_d$ is a measure on the parameter space of the density function $\betadist(t; \alpha, \beta)$; to be precise, we may specify this parameter space as $\Theta \defas \PosReals \times \PosReals$ and equip it with the $\sigma$-algebra of Borel sets $\Bcal(\Theta)$. Then $G_d$ and $H$ are measures on $(\Theta, \Bcal(\Theta))$, and we may specify independent univariate prior distributions on $\alpha$ and $\beta$, such as
\[
H(\alpha, \beta) = \gammadist(\alpha; a_\alpha, b_\alpha) \times \gammadist( \beta; a_\beta, b_\beta )
	,
	\label{eq:basemeasure1}
\]
for some additional hyperparameters $a_\alpha, b_\alpha, a_\beta, b_\beta > 0$.

The Dirichlet process $G_d$ is almost surely finite with a countably infinite number of atoms; we may write
\[
G_d = \sum_{j=1}^\infty \pi_{d,j} \delta_{(\tilde \alpha_j, \tilde \beta_j)}
	,
	\label{eq:boreldp}
\]
for a collection of random elements $\pi_{d,1}, \pi_{d,2}, \dots$ in $(0,1]$ satisfying $\sum_{j=1}^\infty \pi_{d,j} = 1$ \as\ and a collection of random elements $(\tilde \alpha_1, \tilde \beta_1), (\tilde \alpha_2, \tilde \beta_2), \dots$ in $\Theta$.
While there are an infinite number of random variables in \cref{eq:boreldp}, only a finite number of them will be associated with the data and represented during simulation and inference.
To this end, introduce an indicator variable $Z_{d, i}$, for every $i \le N_d$, that denotes the mixture component to which the trade at time $t_{d,i}$ is assigned. That is,
\[
Z_{d,i} \given \pi_d \dist \pi_d
	,
	\qquad
	i \le N_d
	,
\]
where $\pi_d \defas (\pi_{d,1}, \pi_{d,2}, \dots)$.
This sequence may be practically sampled (without representing $G_d$ or $\pi_d$) according to a \emph{Chinese restaurant process} with concentration parameter $\conc_d$.

Let $K$ denote the number of unique labels appearing among $Z_d \defas (Z_{d,1}, \dotsc, Z_{d, N_d})$, and let $(\alpha_k, \beta_k) \in \Theta$, for $k\le K$, denote their corresponding mixture components. Let $\alpha \defas (\alpha_1, \dotsc, \alpha_K)$ and $\beta \defas (\beta_1, \dotsc, \beta_K)$.
Then a practical, generative model for the data on day $d\le D$ may then be summarized as follows:
\begin{enumerate}[noitemsep]
\item Sample $\Lambda_d \dist \gammadist(a_\Lambda, b_\Lambda)$ and the number of trades $N_d \given \Lambda_d \dist \Poisson(\Lambda_d)$.

\item Sample component assignments $Z_{d,1}, \dotsc, Z_{d,N_d}$ from a Chinese restaurant process with concentration parameter $\conc_d>0$. 

\item For every unique component index $k\le K$, sample $(\alpha_k, \beta_k) \dist H$.

\item For every trade $i\le N_d$, sample the trade time
	$$t_{d,i} \given Z_{d,i}, \alpha, \beta \dist \betadist( \alpha_{Z_{d,i}}, \beta_{Z_{d,i}} ).$$
\end{enumerate}
Finally, note that we have the following conditional likelihood
\[
p ( t_d, N_d \given Z_d, \Lambda_d, \alpha, \beta ) 
	= \Lambda_d^{N_d} \exp ( - \Lambda_d ) \prod_{i=1}^{N_d} \betadist ( t_{d,i}; \alpha_{Z_{d,i}}, \beta_{Z_{d,i}} )
	.
\]

\subsection{Modeling trade quantities with marked Poisson processes}
\label{sec:markedpp}

We now follow \citet{taddy2012mixture} to extend $G_d$ to additionally specify the parameters of a distribution on the entire process of cumulative traded volume $X_d(t)$, as defined in the main article.
Note that we are still only considering the trading volume curve for a fixed day $d\le D$, without regard to the trading curves of other days.

We form a marked Poisson process by marking the jumps of the Poisson process $\Ical_d(t)$ with their corresponding trade quantities $q_d \defas (q_{d,1}, \dotsc, q_{d,N_d})$.
Formally, we obtain a (joint) model for $(t_d, q_d)$ by defining a Poisson process on the product space $(0,1) \times \PosInts$ with intensity function $\phi_d$, where
\begin{equation}
\begin{gathered}
\phi_d(t, q; G_d) \defas \Lambda_d \int \betadist(t ; \alpha, \beta) \kappa ( q ; \theta_q ) \dee G_d ( \alpha, \beta, \theta_q )
	,
	\qquad
	t \in (0, 1)
	, \,
	q \in \PosInts
	,
	\\
	G_d \dist \DPLAW(\conc_d, H)
	,
	\qquad
	\Lambda_d \dist \gammadist( a_\Lambda, b_\Lambda )
	,
	\label{eq:jointintensity}
\end{gathered}
\end{equation}
for some probability mass function (\pmf) $\kappa(q ; \theta_q)$ on $\PosInts$ with parameters $\theta_d$, and where $H$ is now a prior distribution on an appropriately extended parameter space. 
We have therefore specified (conditionally) independent models for the trade times $t_d$ and quantities $q_d$, and a dependency between their parameters is induced by the coupling of components under the Dirichlet process.
Note that a conditional likelihood given the data may be written as 
\[
p( t_d, q_d, N_d \given \Lambda_d, G_d )  
	\propto \Lambda_d^{N_d} \exp ( - \Lambda_d )
	 \prod_{i=1}^{N_d} f( t_{d,i}, q_{d,i}; G_d)
	 ,
	 \label{eq:markedpplikel}
\]
where $f(t, q; G_d) = \phi(t, q; G_d) / \Lambda_d$.

To practically sample this process, note that the \emph{marginal intensity function}
\[
\sum\nolimits_{q \in \PosInts} \phi_d(t, q; G_d)
	&= \Lambda_d \int \betadist(t ; \alpha, \beta) 
		\Bigl [ \sum\nolimits_{q \in \PosInts} \kappa ( q ; \theta_q ) \Bigr ] 
		\dee G_d ( \alpha, \beta, \theta_q )
	\\
	&= \lambda_d(t ; G_d)
\]
is the (locally integrable) intensity function in \cref{eq:countintensity}, and it follows from an inverse of the Poisson process marking theorem that $X_d(t)$ is a marked Poisson process with jumps at locations $t_d$ and corresponding marks $q_d$ \citep{taddy2012mixture}.
In particular, we may sample $\Ical_d(t)$ as defined in \cref{sec:timespp}, and conditioned on $Z_{d,i}$ (which we recall denotes the mixture component to which datapoint $t_{d,i}$ is assigned), we may sample the corresponding mark $q_{d,i}$ according to the distribution with \pmf\ $\kappa( q ; \theta_{Z_{d,i}} )$.

To appropriately specify $\kappa(q ; \theta_q)$, recall that the trade quantities, $q_{d,i}$, record the number of lots (of shares) traded at time $t_{d,i}$, and so are positive integers. We could therefore let $\kappa( q ; \theta_q )$ be a zero-truncated negative binomial distribution, for example. In our experiments, however, we instead find that modeling the \emph{additional} number of lots traded (given that at least one lot traded at $t_{d,i}$) is numerically more robust:
\[
\kappa(q_{d,i} ; \theta_q) = \Pr \{ q_{d,i} \given q_{d,i} \ge 1 \} = \nbdist( q_{d,i} - 1 ; r, \tau )
	,
	\qquad
	q_{d,i} \ge 1
	,
\]
where $\theta_q = (r, \tau)$ for some $r, \tau >0$.
Note that we use an uncommon parameterization of the negative binomial distribution, where $\tau = p / (1-p)$ is the odds-ratio of the more commonly used probability of success (or failure) parameter $p\in (0,1)$. The probability mass function is
\[
\nbdist( x ; r, \tau ) = \frac{\Gamma(r + x)}{x! \Gamma(r)} 
	\frac{ \tau^x }{ (\tau + 1)^{r+x} }
	,
	\qquad
	x \in \{ 0 , 1, \dots \}
	.
\]
The base measure $H$ of the Dirichlet process (in \cref{eq:jointintensity}) must be appropriately extended to define a prior distribution over $r$ and $\tau$. We may simply specify independent univariate prior distributions over each parameter, for example, extending the distribution $H$ given in \cref{eq:basemeasure1} to
\[
\label{eq:Hprior}
\begin{split}
H(\alpha, \beta, r, \tau)
 	&= \gammadist(\alpha; a_\alpha, b_\alpha) \times \gammadist( \beta; a_\beta, b_\beta )
		\\
	&\qquad \qquad
	\times \gammadist(r; a_r, b_r) \times \gammadist( \tau; a_\tau, b_\tau )
	.
\end{split}
\]

The joint intensity function in \cref{eq:jointintensity} then becomes
\[
&\phi_d(t, q ; G_d) 
	\nonumber
	\\
	&\quad
	= \Lambda_d \int \betadist( t; \alpha, \beta ) 
	\nbdist( q - 1 ; r, \tau )
	\dee G_d ( \alpha, \beta, r, \tau )
	,
	\quad
	t \in (0, 1)
	, \,
	q \in \PosInts
	,
	\label{eq:finaljointintensity}
\]
and we have the following conditional likelihood given the data
\[
\begin{split}
&p ( t_d, q_d, N_d \given Z_d, \Lambda_d, \alpha, \beta, r, \tau ) 
	\\
	&\qquad \qquad
	\propto \Lambda_d^{N_d} \exp ( - \Lambda_d )
	 \prod_{i=1}^{N_d} \Bigl [ 
	 	\betadist( t_{d,i} ; \alpha_{Z_{d,i}}, \beta_{Z_{d,i}} ) \nbdist( q_{d,i} - 1 ; r_{Z_{d,i}}, \tau_{Z_{d,i}} ) 
	\Bigr ]
	,
\end{split}
\]
where $r\defas (r_1, \dotsc, r_K)$ and $\tau \defas (\tau_1, \dotsc, \tau_K)$, recalling that $K$ represents the number of unique labels among the $Z_d$.
We have now completed our model description for $X_d(t)$, and the generative model summarized at the end of \cref{sec:markedpp} becomes:
\begin{enumerate}[noitemsep]
\item Sample $\Lambda_d \dist \gammadist(a_\Lambda, b_\Lambda)$ and the number of trades $N_d \given \Lambda_d \dist \Poisson(\Lambda_d)$.

\item Sample component assignments $Z_{d,1}, \dotsc, Z_{d,N_d}$ according to a Chinese restaurant process with concentration parameter $\conc_d>0$. 

\item For every unique component index $k\le K$, sample $(\alpha_k, \beta_k, r_k, \tau_k) \dist H$.

\item For every trade $i\le N_d$, sample the trade time
	$$t_{d,i} \given Z_{d,i}, \alpha, \beta \dist \betadist( \alpha_{Z_{d,i}}, \beta_{Z_{d,i}} )$$
and trade quantity $q_{d,i} \defas q_{d,i}' + 1$, where
	$$q_{d,i}' \given Z_{d,i}, r, \tau \dist \nbdist( r_{Z_{d,i}}, \tau_{Z_{d,i}} ).$$
\end{enumerate}

\subsection{Coupling across trading days with the hierarchical Poisson process}
\label{sec:hpp}

We have so far only modeled the trading volume curve for a single day in the dataset depicted in \cref{fig:trade-data}. We now consider jointly modeling all days in the dataset by coupling the Dirichlet processes (determining the Poisson process intensity functions) with a hierarchical construction. 
In particular, we would like to share information (i.e., structure) between the curves $X_1(t), \dotsc, X_D(t)$. Note that we may view the corresponding trade times and quantities for each day $\{(t_d, q_d) \colon d\le D\}$ as $D$ \emph{groups} of data.
A natural way to model such \emph{admixtures} is with the \emph{hierarchical Dirichlet process} \citep{tehhdp}.

The model is as follows:
Let
\[
G_0 &\dist \DPLAW(\conc_0, H)
	\\
G_d \given G_0 &\dist \DPLAW(\conc_d, G_0)
	,
	\qquad
	d\le D
	,
	\label{eq:hdp}
\]
where $\conc_0 > 0$ is an additional \emph{global concentration parameter} and, as before, $H$ may be interpreted as a prior distribution on the component parameters.
Constructing the intensity function $\lambda_d (t; G_d)$ in \cref{eq:countintensity} for each day $d\le D$ with the coupled Dirichlet processes $G_d$ completes the description of our model.
We call this model for the collection $\Ical_1(t), \dotsc, \Ical_D(t)$, characterized by the intensity functions $\lambda_1(t; G_1), \dotsc, \lambda_D(t,; G_D)$, a \emph{hierarchical Poisson process}, or correspondingly we may call the model for $X_1(t), \dotsc, X_D(t)$ a \emph{hierarchical marked Poisson process}.

To be practical, we may once again introduce $Z_{d,i} \given G_d \dist G_d$ to denote the mixture component associated with datapoint $(t_{d,i}, q_{d,i})$, for every $d\le D$ and $i\le N_d$.
In this case, the collection $Z \defas (Z_1, \dotsc, Z_d)$ may be sampled according to the \emph{Chinese restaurant franchise} described by \citet{tehhdp}. We do not detail this algorithm here for the sake of clarity and space, but its interpretation will be necessary for the description of inference later. 
Let $K$ denote the number of unique values among the elements of $Z$, and let $(\alpha_k, \beta_k, r_k, \tau_k)$, $k\le K$, denote their corresponding \emph{global} mixture components.
A conditional likelihood given the dataset may then be written as
\[
\begin{split}
p(t, q, N \given Z, \Lambda) 
	&\propto
	\prod_{d\le D} \Bigl [
	\Lambda_d^{N_d} 
	\exp ( - \Lambda_d ) 
		\\
		&\qquad \times
	\prod_{i=1}^{N_d} 
		\bigl [ 
			\betadist( t_{d,i}; \alpha_{Z_{d,i}}, \beta_{Z_{d,i}} ) 
			\nbdist( q_{d,i} - 1 ; r_{Z_{d,i}}, \tau_{Z_{d,i}} )
		\bigr ]
		\Bigr ]
	,
\end{split}
\]
where $t \defas (t_1, \dotsc, t_D)$, $q \defas (q_1, \dotsc, q_D)$, 
$N\defas (N_1, \dotsc, N_D)$, and $\Lambda \defas (\Lambda_1, \dotsc, \Lambda_D)$.

We conclude this section by noting that our hierarchy may be straightforwardly extended to naturally capture broader groupings of the dataset, for example, across different stocks. Furthermore, financial instruments are sensibly grouped at even coarser levels, for example, into small-, medium-, and large-cap stocks, or grouped by the sector of the company (industrial, food and beverage, financial services, etc.). 
An example of the hierarchical Poisson process construction for the latter example could look as follows:
\[
&\text{Global level:}
&&G_0 \dist \DPLAW(\conc_0, H)
	,
	\\
&\text{Sector level:}
&&G_s \given G_0 \dist \DPLAW(\conc_s, G_0)
	,
	\qquad s \le S ,
	\\
&\text{Stock level:}
&&G_{s,j} \given G_s \dist \DPLAW(\conc_{s,j} , G_s ) 
	,
	\qquad s\le S, \, j\le J_s ,
	\\
&\text{Day level:}
&&G_{s,j,d} \given G_{s,j} \dist \DPLAW(\conc_{s,j,d}, G_{s,j})
	,
	\qquad s \le S, \, j\le J_s, \, d\le D_j ,
\]
where $S$ is the number of sectors, $J_s$ is the number of stocks in sector $s$, and $D_j$ is the number of days recorded for stock $j$.
The times $(t_{s,j,d})$ and quantities $(q_{s,j,d})$ 
for the trades on day $d$ of stock $j$ in sector $s$ are sampled from a Poisson process with intensity function
\[
&\phi_{s,j,d} (t, q ; G_{s,j,d})
	\nonumber
	\\
	&\quad
	= \Lambda_d \int \betadist( t; \alpha, \beta ) 
	\nbdist( q - 1 ; r, \tau )
	\dee G_{s,j,d} ( \alpha, \beta, r, \tau )
	,
	\quad
	t \in (0, 1)
	, \,
	q \in \PosInts
	.
\]

\section{Posterior inference}
\label{sec:inference}

To develop a computationally efficient Markov Chain Monte Carlo (MCMC) sampler for the hierarchical Poisson process, we will rely on a stick-breaking representation for the hierarchical Dirichlet process due to \citet{tehhdp} and the now-popular slice sampling approaches for Dirichlet process mixture models \citep{walker2007sampling}.
In particular, we say that the sequence of Dirichlet process weights $\nu$, constructed according to the stick-breaking representation in \cref{eq:localsticks}, has a \emph{GEM distribution with concentration parameter $\conc$}, and we write $\nu \dist \GEMLAW(\conc)$. 
(The letters stand for Griffiths--Engen--McCloskey \citep{pitman2006combinatorial}.)
Then consider the alternative (equivalent) representation of the hierarchical Poisson process model
\[
&\nu \dist \GEMLAW(\conc_0)
	\\
&\pi_d \dist \GEMLAW(\conc_d)
	,
	&&\qquad d\le D
	,
	\\
&K_{d,t} \given \nu \dist \nu
	,
	&&\qquad d\le D, \, t = 1,2,\dotsc
	,
	\\
&T_{d,i} \given \pi_d \dist \pi_d
	,
	&&\qquad d\le D, \, i\le N_d
	,
	\\
&t_{d,i}, q_{d,i} \given K_d, T_{d,i} \dist F(\theta_{K_{d,T_{d,i}}})
	,
	&&\qquad d\le D, \, i\le N_d
	,
\]
where $F(\theta_k)$ is the distribution with density function $f(t, q ; \theta_k)$ with parameter $\theta_k$ described in \cref{sec:timespp}, and $\theta_1, \theta_2, \dots$ are distributed \iid\ according to $H$ (given by \cref{eq:basemeasure1}).
\citet{tehhdp} showed that this construction is equivalent to the construction in \cref{sec:timespp}, where the assignment labels are given by $Z_{d,i} = K_{d,T_{d,i}}$. 
In the Chinese restaurant franchise described therein, $T_{d,i}$ denotes the table that customer $i$ sits at in restaurant $d$, and $K_{d,t}$ denotes the dish served at table $t$ in restaurant $d$. We do not review this metaphor here, but we do refer to it throughout the rest of the section and direct the unfamiliar reader to \citet{tehhdp} for this background.

It is straightforward to develop a slice sampler for this representation of the hierarchical Dirichlet process by applying the techniques introduced by \citet{walker2007sampling,kalli2011slice}, as applied, for example, by \citet{amini2019exact}. 
Here, we present a novel variant of this algorithm with several properties that are important for computational efficiency.
For simplicity, focus on the following conditional likelihood:
\[
\begin{split}
p( t, q, K, T, \nu, \pi, \given N, \theta )
	&=
	\prod_{d=1} \Biggl [ 
	\prod_{i=1}^{N_d} \Bigl [ 
		f( t_{d,i}, q_{d,i}; \theta_{K_{d,T_{d,i}}} ) \pi_{d,T_{d,i}}
		\Bigr ]
	\prod_{t=1}^\infty \nu_{K_{d,t}} 
	\Biggr ]
	\\
	&\qquad \qquad \times
	\prod_{k=1}^\infty
		\betadist( \nu_k' ; 1 , \conc_0 )
	\prod_{t=1}^\infty 
		\betadist( \pi_{d,t}' ; 1, \conc_d )	
	.
\end{split}
\label{eq:marginalslice}
\]
The general idea is to introduce collections of \emph{auxiliary slice variables}, conditioned on which, the number of mixture components that need to be represented during inference will be finite. To this end, let
\[
u_{d,i} \given \pi_d, T_{d,i} 
	\dist \Uniform( 0 , \pi_{d,T_{d,i}} )
	,
	\qquad 
	d\le D, \, i\le N_d
	,
\]
where $\pi_d \defas (\pi_{d,1}, \pi_{d,2}, \dots)$, for every $d\le D$, and let
\[
w_{d,t} \given \nu, K_d
	\dist \Uniform( 0 , \nu_{K_{d,t}} )
	,
	\qquad
	d\le D, \, t = 1,2,\dotsc
	,
\]
where $\nu \defas (\nu_1, \nu_2, \dots)$.
Then we have 
\[
\begin{split}
&p( t, q, K, T, \nu, \pi, u, w \given N, \theta )
	\\
	&\qquad =
	\prod_{d=1} \Biggl [ 
	\prod_{i=1}^{N_d} \Bigl [ 
		f( X_{d,i}; \theta_{K_{d,T_{d,i}}} ) \Indicator{ u_{d,i} < \pi_{d,T_{d,i}} }
		\Bigr ]
	\prod_{t=1}^\infty \Indicator{ w_{d,t} < \nu_{K_{d,t}} }
	\Biggr ]
	\\
	&\qquad \qquad \times
	\prod_{k=1}^\infty
		\betadist( \nu_k' ; 1 , \conc_0 )
	\prod_{t=1}^\infty 
		\betadist( \pi_{d,t}' ; 1, \conc_d )	
	,
\end{split}
\label{eq:jointslice}
\]
where $u \defas (u_{d,i}, d\le D, \, i\le N_d)$ and $w \defas (w_{d,t}, d\le D, \, t=1,2,\dots)$.
Note that integrating the ``joint'' likelihood in \cref{eq:jointslice} over $u$ and $w$ simply reduces back to the ``marginal'' likelihood in \cref{eq:marginalslice}. 
Note then that the auxiliary uniform variates $u$ and $w$ play the role of \emph{horizontal slices} in the sense of a \emph{slice sampling} algorithm \citep{neal2003slice,robert2013monte}. In particular, the presence of the indicator functions in \cref{eq:jointslice} will act to truncate the support of all conditional distributions of interest to finite sets. 
Importantly, this algorithm allows us to parallelize resampling of the indicator variables $K_{d,t}$ and $T_{d,i}$; scalability and efficient Markov chain mixing will depend particularly on the latter. This is to be contrasted to Gibbs-sampling style MCMC procedures, usually based on the Chinese restaurant franchise representations derived by \citet{tehhdp}, which iterate such resampling steps over every datapoint.

The main steps of the MCMC algorithm are summarized by the conditional distributions below. Most steps are straightforward to derive from \cref{eq:jointslice} (and as described by \citet{walker2007sampling,kalli2011slice}), with some notable (and important) exceptions that will be detailed later.
%


{\bf Sample $\nu$}: 
Sample the weights $\nu_1, \dotsc, \nu_{K^*}$ of the \emph{global} Dirichlet process associated with the $K^*$ occupied components, along with an additional variable $\nu^*$, according to
\[
( \nu_1, \dotsc, \nu_{K^*}, \nu^* ) \given T^*
	\dist \Dirichlet ( m_1, \dotsc, m_{K^*} , \conc_0 )
	,
\]
where $T^* \defas (T_1^*, \dotsc, T_D^*)$ and
\[
m_k \defas \sum_{d=1}^D \sum_{t=1}^{T_d^*} 
	\Indicator{K_{d,t} = k}
	,
	\qquad k = 1, \dotsc, K^*
	.
	\label{eq:tabledishcounts}
\]
Note that $\nu^* = 1 - \sum_{k=1}^{K^*} \nu_k$.


{\bf Sample $\pi_d$}:
For every $d\le D$, sample the weights $\pi_{d,1} ,\dotsc, \pi_{d,T_d^*}$ of the \emph{local} Dirichlet process associated with the $T_d^*$ occupied tables (in restaurant $d$), along with an additional variable $\pi_d^*$, according to
\[
( \pi_{d,1} ,\dotsc, \pi_{d,T_d^*}, \pi_d^* ) \given T_d
	\dist \Dirichlet ( n_{d,1}, \dotsc, n_{d,T_d^*} , \conc_d )
	,
\]
where
\[
n_{d,t} \defas \sum_{i=1}^{N_d} \Indicator{ T_{d,i} = t } 
	,
	\qquad
	t = 1, \dotsc, T_d^*
	.
\]
Note that $\pi_d^* = 1 - \sum_{t=1}^{T_d^*} \pi_{d,t}$.


{\bf Sample $u_{d,i}$}: 
For every $d\le D$, sample the slice variables associated with the datapoints
\[
u_{d,i} \given T_{d,i} \dist \Uniform(0, \pi_{d,T_{d,i}})
	,
	\qquad i = 1,\dotsc, N_d
	,
\]
and set $u_d^* \defas \min \{ u_{d,i} \colon i\le N_d \}$.


{\bf Sample new local tables}:
After resampling the slice variables $u$, we may need to represent additional tables. For every $d\le D$, do the following.
\begin{enumerate}
\item If $\pi_d^* < u_d^*$, then we don't need to represent additional tables in this restaurant. \label{step:newcomp}

\item Otherwise, iteratively add new tables to restaurant $d$ as follows until $\pi_d^* > u_d^*$.

\begin{enumerate}

\item Sample the DP weight $\pi_{d, T_d^* + 1}$ to correspond to the new table and redefine the remaining stick length $\pi_d^*$ according to
\[
b \dist \betadist ( 1, \conc_d ) 
	,
	\qquad
	\pi_{d,T_d^* + 1} = b \pi_d^*
	,
	\qquad
	\pi_d^* = (1-b) \pi_d^*
	.
\]

\item Sample a dish to serve at the table according to the CRF prior, in particular, we treat the table $T_{d,T_d^* + 1}$ as the last table (amongst all restaurants) to be sampled in the process, that is,
\[
\Pr \{ K_{d,T_d^*+ 1} = k \given K_d \}
	\propto
	m_k \Indicator{ k \le K^* }
	+ 
	\conc_d \Indicator{ k = K^* + 1 }
	.
\]
where $m_k$ is the number of currently represented tables (across all restaurants) serving dish $k$, which is more precisely given in \cref{eq:tabledishcounts}.
If a previously unrepresented dish is selected to be served at the table, i.e., $K_{d,T_d^*+ 1} = K^* + 1$, then:

\begin{enumerate}
\item Sample a parameter $\theta_{K^* + 1} \dist H$ to associate with the new dish.

\item Sample a global DP weight $\nu_{K^*+1}$ to correspond to the new dish and redefine the remaining global DP stick length $\nu^*$ according to
\[
b \dist \betadist ( 1, \conc_0 ) 
	,
	\qquad
	\nu_{K^* + 1} = b \nu^* 
	,
	\qquad
	\nu^* = (1-b) \nu^*
	.
	\label{eq:newglobalstick}
\]

\item Redefine $K^* = K^* + 1$.
\end{enumerate}
	
\item Sample a slice variable for the new table $w_{d,T_d^* + 1} \given K_{d, T_d^* + 1} \dist \Uniform(0, \nu_{K_{d,T_d^* + 1}})$.

\item Redefine $T_d^* = T_d^* + 1$.
\end{enumerate}

\end{enumerate}


{\bf Sample $w_{d,t}$}:
For every $d\le D$, sample the slices
\[
w_{d,t} \given \nu, K_d \dist \Uniform( 0, \nu_{K_{d,t}} )
	,
	\qquad
	t = 1,\dotsc, T_d^*
	,
\]
and set $w^* \defas \min \{ w_{d,t} \colon d\le D,\, t \le T_d^* \}$.


{\bf Sample new global dishes}:
After resampling the slice variables $w$, we may need to increase the number of represented global dishes. 
\begin{enumerate}
\item If $\nu^* < w^*$, then we don't need to sample new dishes.  \label{step:globalnewcomp}

\item Otherwise, iteratively sample new global dishes as follows until $\nu^* > w^*$.

\begin{enumerate}
\item Sample new component parameters $\theta_{K^* + 1} \dist H$ to associate with the new dish.

\item Sample the new global DP weight $\nu_{K^* + 1}$ and redefine the remaining global stick length $\nu^*$ according to \cref{eq:newglobalstick}.

\item Set $K^* = K^* + 1$.
\end{enumerate}

\end{enumerate}


{\bf Sample $K_{d,t}$}:
For every $d\le D$ and $t \le T_d^*$, sample $K_{d,t}$ from the conditional distribution
\[
\begin{split}
&\Pr \{ K_{d,t} = k \given X_d, T_d, w_{d,t}, \nu_k, \theta_k \}
	\\
	&\qquad \propto
	\begin{cases}
	\prod_{ \{i \le N_d \colon T_{d,i} = t \} }
		f( X_{d,i} ; \theta_k )
		,
	&\quad k \in \{ k' \le K^* \colon \nu_{k'} > w_{d,t}  \}
	,
	\\
	0 , &\quad \text{otherwise.}
	\end{cases}
\end{split}
\]
Note that, after resampling all indicators $K_{d,t}$, any ``empty'' global components (i.e., dishes not being served at any tables in any restaurant) need to be dropped.


{\bf Sample $T_{d,i}$}:
For every $d\le D$ and $i\le N_d$, sample $T_{d,i}$ from the conditional distribution
\[
\Pr\{ T_{d,i} = t \given u_{d,i}, K_d, \pi_d, \theta \}
	\propto
	\begin{cases}
	f( X_{d,i} ; \theta_{K_{d, t}} )
	,
	&\qquad
	t \in \{ t' \le T_d^* \colon \pi_{d,t'} > u_{d,i}  \}
	\\
	0 , &\qquad \text{otherwise.}
	\end{cases}
\]
Note that, after resampling all indicators $T_{d,i}$, any empty local components (i.e., tables with no customers sitting at them) need to be dropped.

{\bf Sample $\theta_k$}:
For every instantiated component $k\le K^*$, resample the component parameters $\theta_k$ from the conditional distribution
\[
p( \theta_k \given X, T, K ) 
	\propto h( \theta_k ) 
	\prod_{(d,i) \in \Ical_k } 
	f ( t_{d,i}, q_{d,i} ; \theta_k )
	,
\]
where $h$ is the density function of $H$, and the product is over the set of all customers (across all restaurants) eating dish $k$, i.e.,
\[
\Ical_k \defas \{ (d,i) \colon d \le D, \, i\le N_d ,\, K_{d,T_{d,i}} = k \}
	,
	\qquad k \le K^*
	,
	\label{eq:paramposterior}
\]
which could be empty, in which case $\theta_k$ is simply resampled from its prior $H$. In our experiments, we sample from \cref{eq:paramposterior} using slice sampling \citep{neal2003slice}.

We take an ``epoch-based'' approach to the MCMC sampler, where for every $d\le D$, we resample all indicators (simultaneously) for fixed $d$ followed by resampling passes on all parameters. The entire algorithm is summarized in \cref{alg:inference}.

\begin{algorithm}
\small
\begin{enumerate}
\item Initialization.
\begin{enumerate}[nolistsep,noitemsep]
\item For every $d\le D$, initialize $T_d \defas (T_{d,1}, \dotsc, T_{d,N_d})$ and $K_d \defas (K_{d,1}, \dotsc, K_{d,T_d^*})$ according to the Chinese restaurant franchise, where $T_d^*$ is the number of tables initialized in restaurant $d$, i.e., the number of unique labels among $T_d$.

\item Set $K^*$ to be the number of instantiated dishes, i.e., the number of unique labels appearing among all $K_{d,t}$, $d\le D$, $t\le T_d^*$.

\item Sample $\theta_k \dist H$, where $\theta_k \defas (\alpha_k, \beta_k, r_k, \tau_k)$, \iid\ for every $k\le K^*$.
\end{enumerate}

\item Iterate the following steps for every $d \le D$ repeatedly until you are satisfied your Markov chain has mixed.

\begin{enumerate}
\item Sample stick-breaking weights $\nu$ of the global Dirichlet process.

\item Sample the slice variables $W_d$.

\item Potentially create new global components (i.e., dishes); if so, resample $\nu$ again after the new components are instantiated.

\item Sample all indicators in $K_d$ simultaneously, then drop any empty global components. 

\item Sample the local stick-breaking weights $\pi_d$.

\item Sample the slice variables $U_d$.

\item Potentially create new local components (i.e., tables in restaurant $d$); if so, resamples $\pi_d$ again after the new components are instantiated.

\item Sample all indicators in $T_d$ simultaneously, then drop any empty local components.

\item Sample component parameters $\theta_k$, $k\le K^*$, concentration parameters $\gamma_0$ and $\gamma_d$, $d\le D$, and any hyperparameters.

\end{enumerate}
\end{enumerate}
\caption{MCMC inference algorithm}
\label{alg:inference}
\end{algorithm}

\section{Results}
\label{sec:experiments}

\begin{figure}[th!]
\centering
\subfloat[AAPL - mixture components]{
	\includegraphics[scale=0.45]{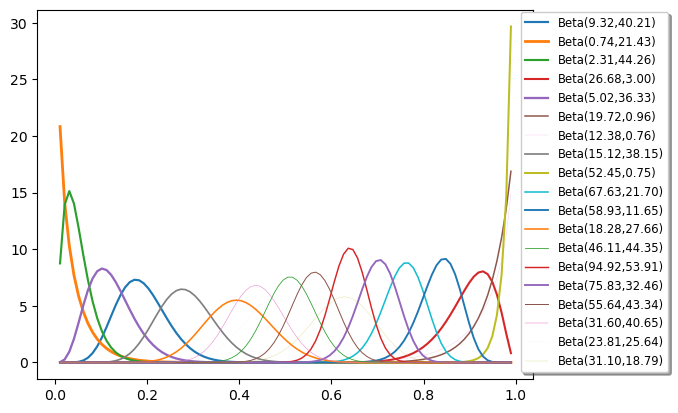}
	\label{fig:aapl_components}
}
\subfloat[AAPL - intensity function]{
	\includegraphics[scale=0.45]{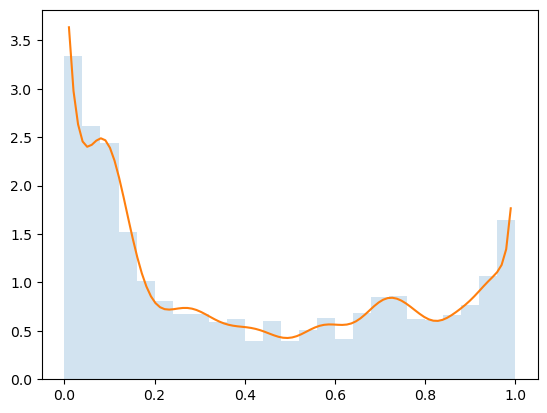}
	\label{fig:aapl_intensity}
}\\
\subfloat[IRBT - mixture components]{
	\includegraphics[scale=0.45]{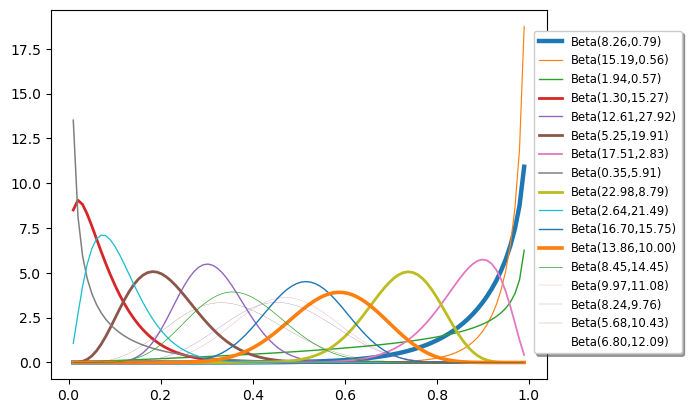}
	\label{fig:irbt_components}
}
\subfloat[IRBT - intensity function]{
	\includegraphics[scale=0.45]{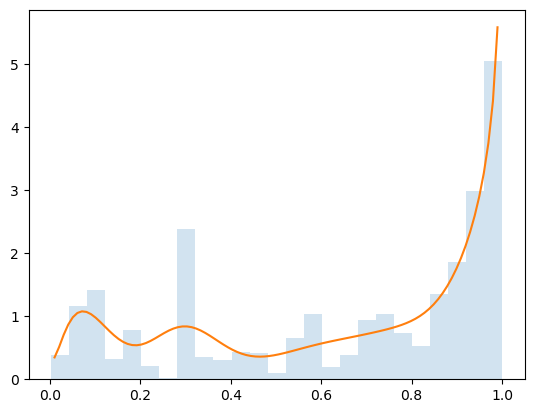}
	\label{fig:irbt_intensity}
}\\
\subfloat[PZZA - mixture components]{
	\includegraphics[scale=0.45]{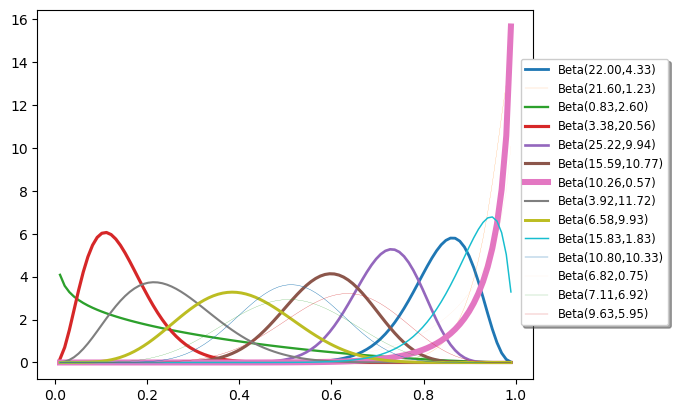}
	\label{fig:pzza_components}
}
\subfloat[PZZA - intensity function]{
	\includegraphics[scale=0.45]{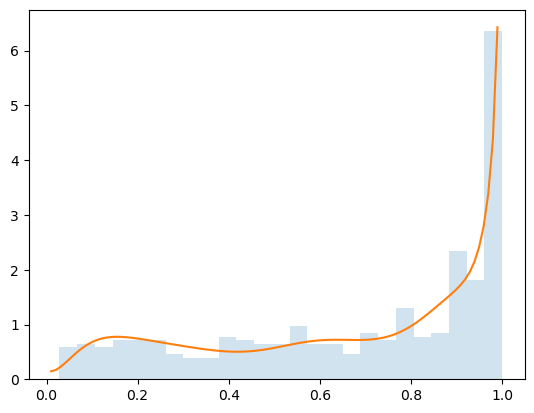}
	\label{fig:pzza_intensity}
}
\caption{The mixture components and intensity functions inferred for one day of a volume trading curve (for the MAP sample from among 1,000 MCMC samples following the burn-in period) for each stock.}
\label{fig:results}
\end{figure}

\begin{figure}[th!]
\centering
\subfloat[Global mixture components ($K^*$)]{
	\includegraphics[scale=0.2]{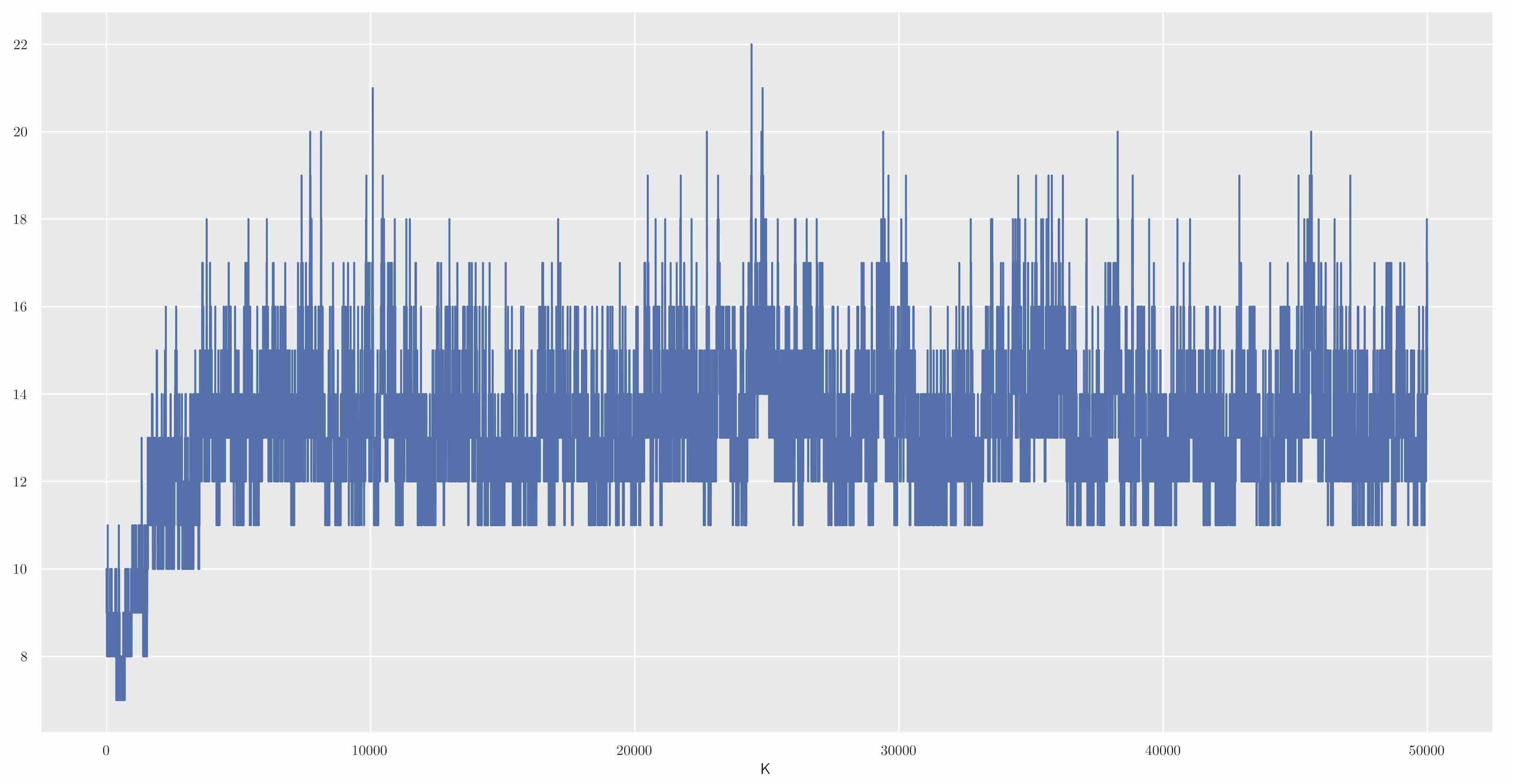}
	\label{fig:pzza_K_trace}
}\\
\subfloat[Mixture components by group ($T_d^*$)]{
	\includegraphics[scale=0.2]{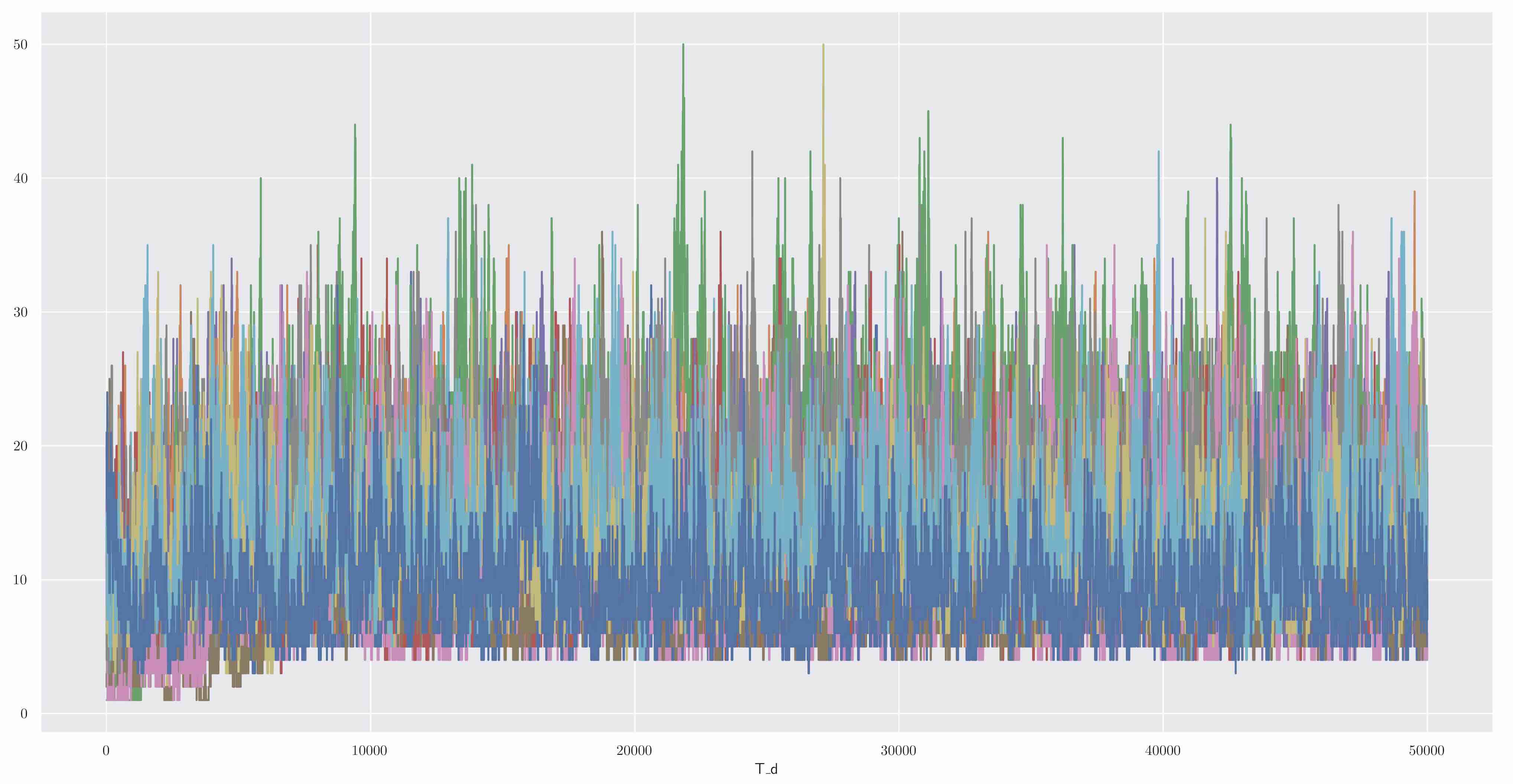}
	\label{fig:pzza_Td_trace}
}
\caption{Trace plots of the number of global mixture components $K^*$ and the number of mixture components $T_d^*$ in each admixture group $d\le D$ over inference epochs for PZZA.}
\label{fig:traceplots}
\end{figure}

\begin{figure}[th!]
\centering
\subfloat[AAPL $K^*$]{
	\includegraphics[scale=0.5]{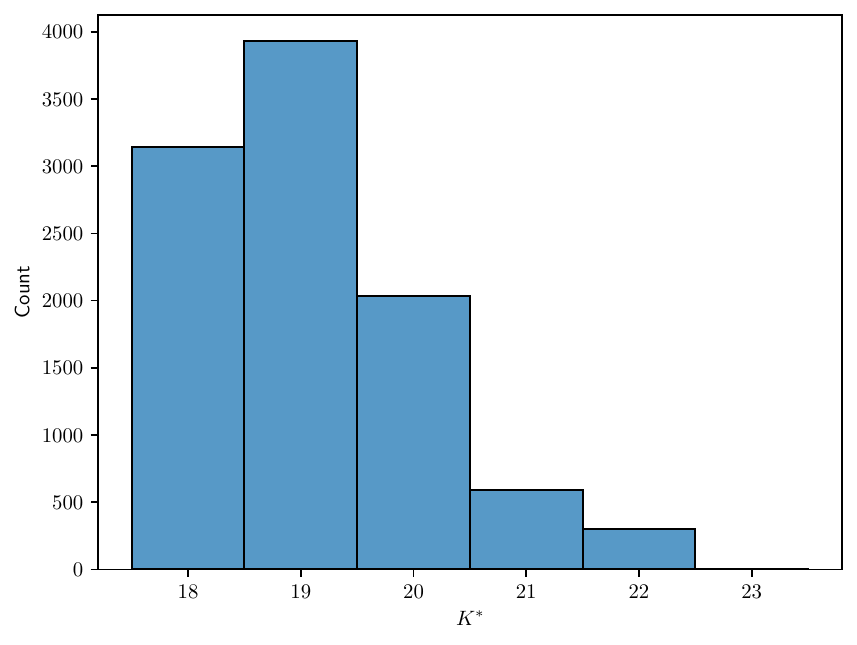}
	\label{fig:K_aapl}
}
\subfloat[AAPL $T_d^*$]{
	\includegraphics[scale=0.5]{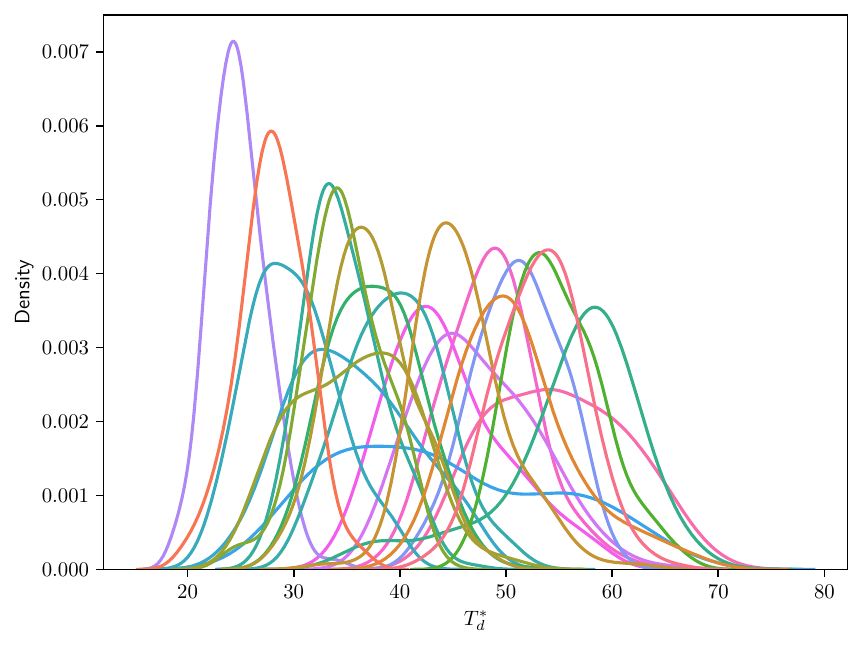}
	\label{fig:Td_aapl}
}\\
\subfloat[IRBT $K^*$]{
	\includegraphics[scale=0.5]{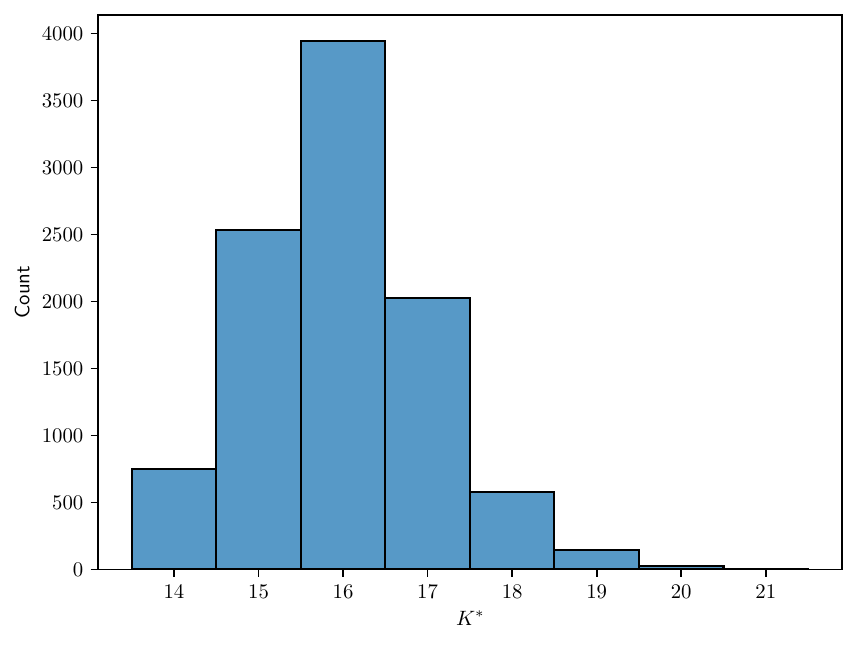}
	\label{fig:K_irbt}
}
\subfloat[IRBT $T_d^*$]{
	\includegraphics[scale=0.5]{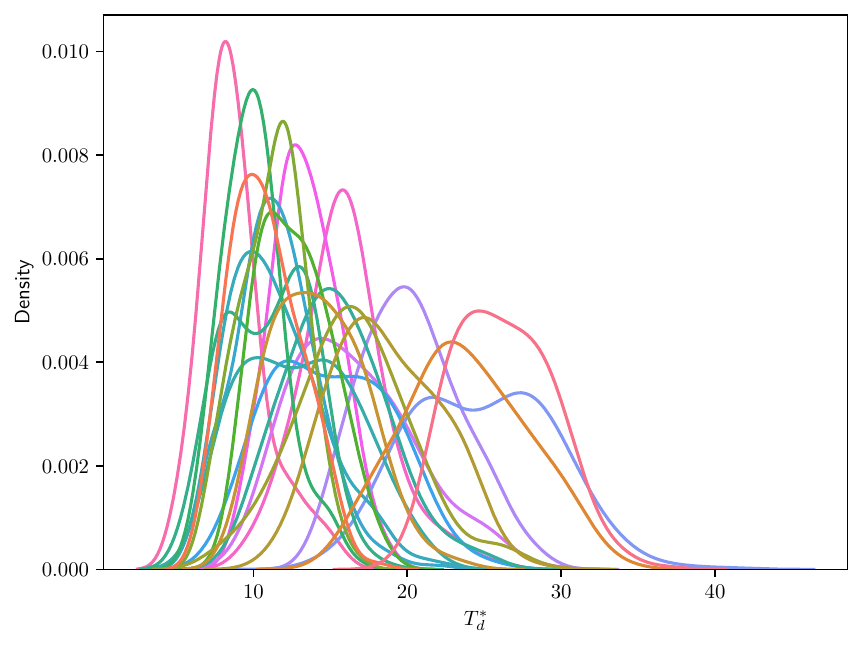}
	\label{fig:Td_irbt}
}\\
\subfloat[PZZA $K^*$]{
	\includegraphics[scale=0.5]{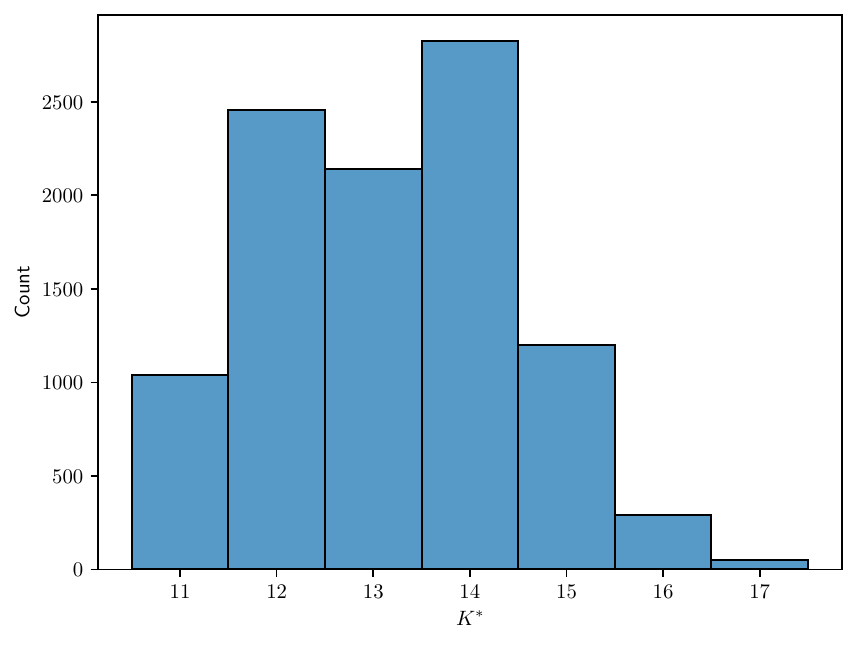}
	\label{fig:K_pzza}
}
\subfloat[PZZA $T_d^*$]{
	\includegraphics[scale=0.5]{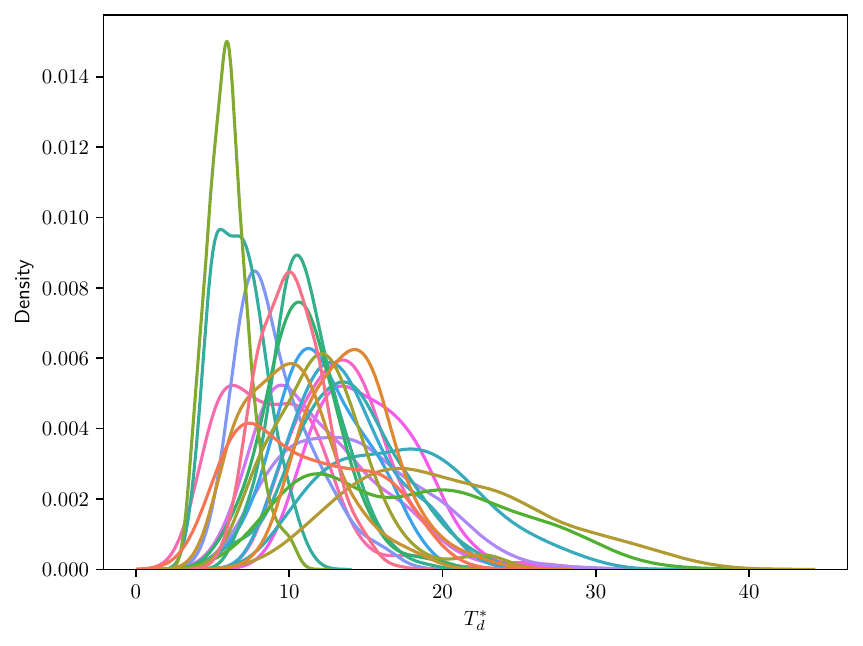}
	\label{fig:Td_pzza}
}
\caption{Visualizations of the inferred distributions over the number of global and local mixture components.}
\label{fig:K_and_Td}
\end{figure}

In \cref{fig:results}, we visualize the mixture components inferred for each stock. The components are displayed for one particular day (arbitrarily chosen) using the maximum a posteriori (MAP) sample from among 1,000 samples taken from the Markov chain following the burn-in period. The mixture components are beta density functions and the two shape parameters are also given in the figure. The weight of the line for each mixture component is proportional to its popularity amongst the admixture groups (i.e., days). In particular, for this given sample, the weight of the line is determined by the relative weight in $\pi_d$. 
The corresponding intensity function for this day is also shown in the figure, overlayed with a histogram of the trading times.

We see that AAPL, with its large number of trades produces a rather smooth histogram with corresponding inferred intensity function. IRBT, as a medium-cap stock, has fewer trades and a less smooth histogram. The resulting intensity function does not hug the histogram tightly. In general, we have found that the intensity function will not fit these trading time histograms on any given day too closely if there is not a large amount of data -- this is a feature of model smoothing in Bayesian inference. In the case of PZZA, despite have the smallest numbers of trades, the histograms are in general smoother than IRBT (trades are spaced more evenly throughout the day and not clustered), and so the inferred intensity functions are still smooth and do not exhibit the behavior seen with IRBT.
We also see that the weighting on mixture components in PZZA is unequal, where one component dominates the mixture weighting distribution. IRBT appears slightly less skewed, and the mixture distribution for AAPL appears somewhat evenly distributed across several components.

To get a sense of inference over the number of mixture components, we display in \cref{fig:traceplots} trace plots of the number of global mixture components $K^*$ and number of local mixture components $T_d^*$, for each day $d\le D$, for PZZA over the inference epochs. In \cref{fig:K_and_Td}, we display distributions over $K^*$ (as histograms) and $T_d^*$ (as kernel density estimates) for each stock generated by 10,000 MCMC samples collected following burn in. 
AAPL has the most ``homogenous'' behavior across different trading days, in terms of the complexity (number of mixture components) of the intensity function. IRBT and PZZA have more heterogeneity, where some trading days are more complex than others. IRBT appears to have the most diversity across the trading days.

\begin{figure}[th!]
\centering
\subfloat[AAPL $\alpha$]{
	\includegraphics[scale=0.5]{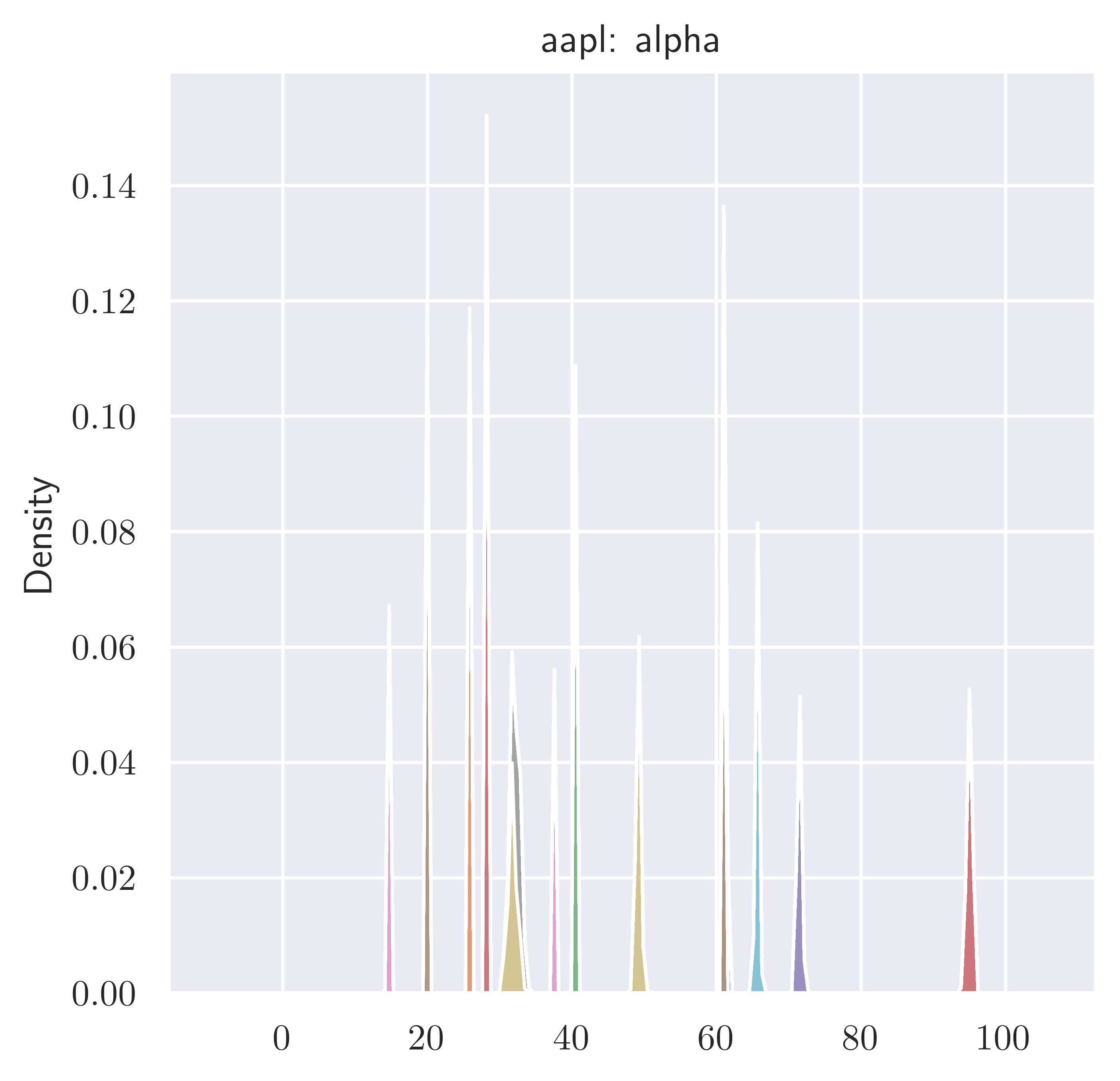}
	\label{fig:aapl_alpha}
}
\subfloat[AAPL $\beta$]{
	\includegraphics[scale=0.5]{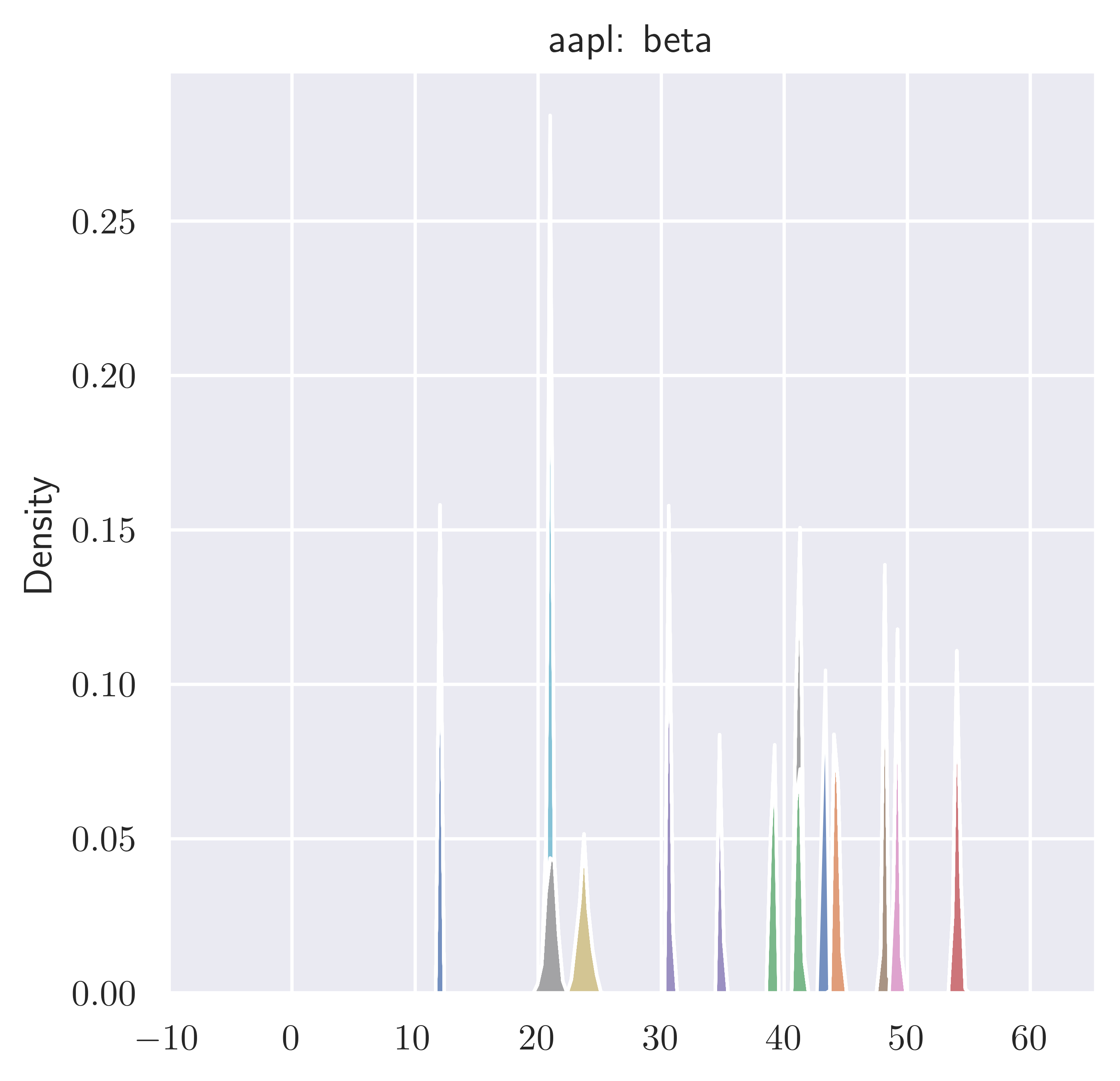}
	\label{fig:aapl_beta}
}\\
\subfloat[IRBT $\alpha$]{
	\includegraphics[scale=0.5]{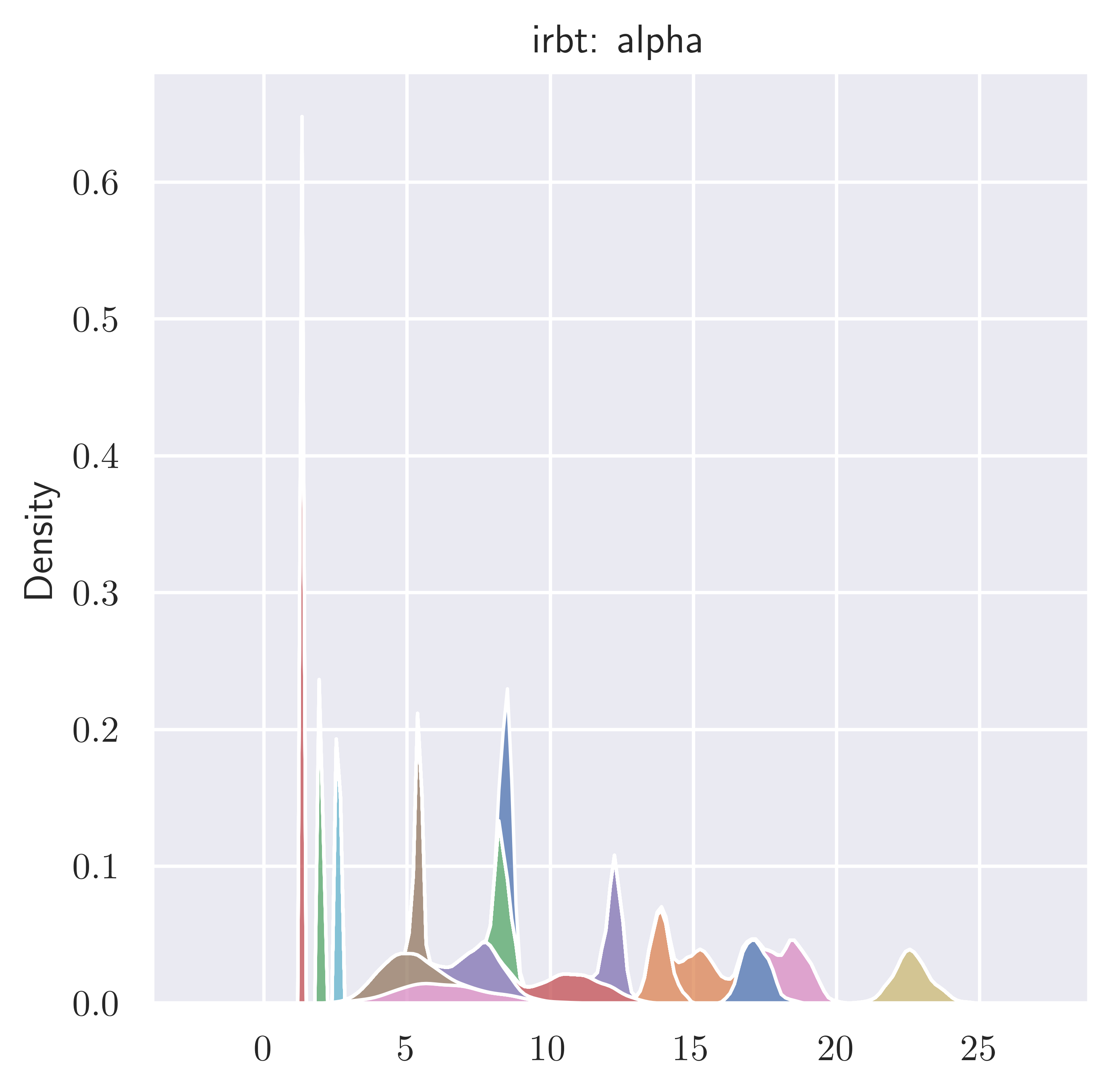}
	\label{fig:irbt_alpha}
}
\subfloat[IRBT $\beta$]{
	\includegraphics[scale=0.5]{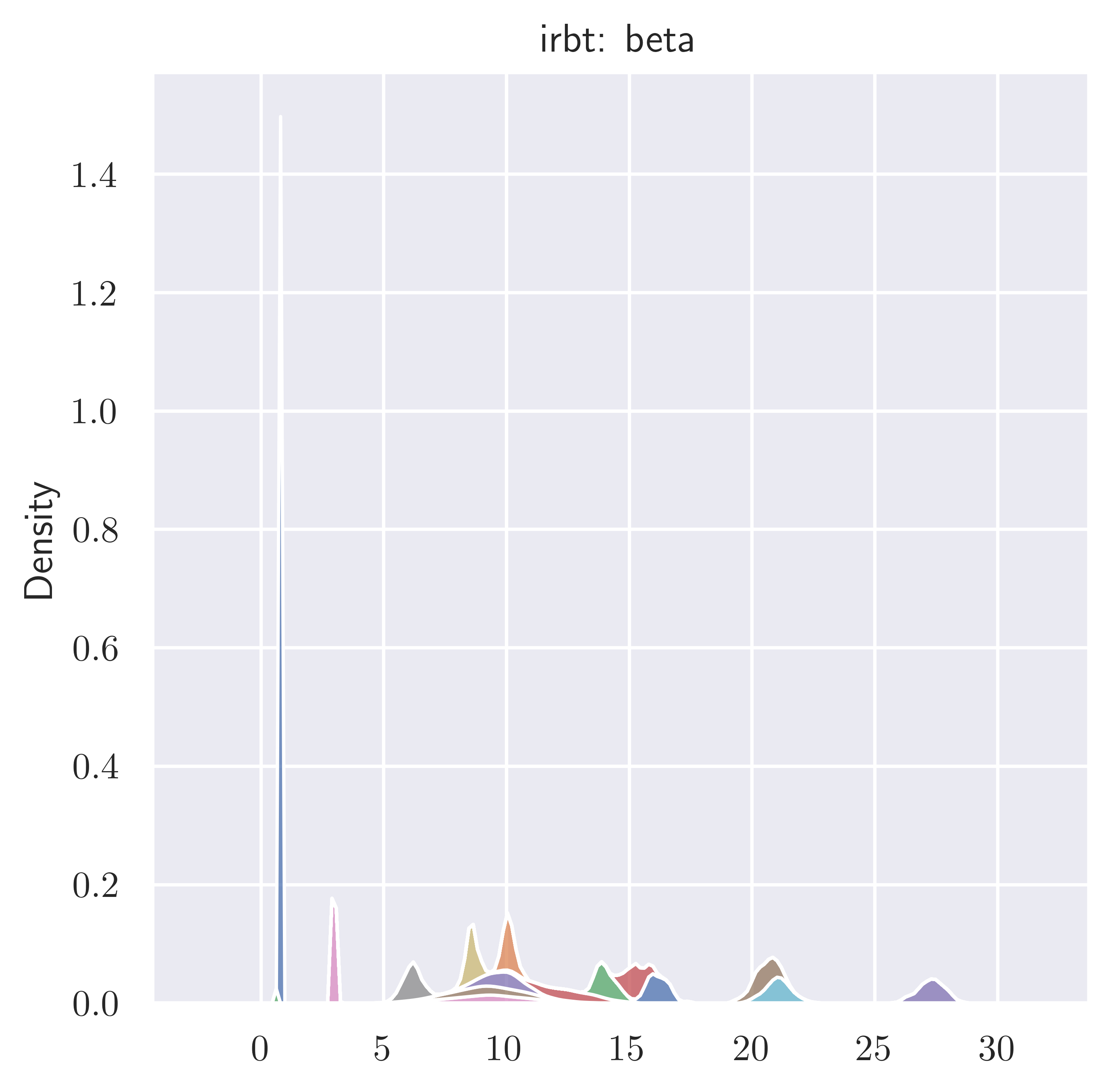}
	\label{fig:irbt_beta}
}\\
\subfloat[PZZA $\alpha$]{
	\includegraphics[scale=0.5]{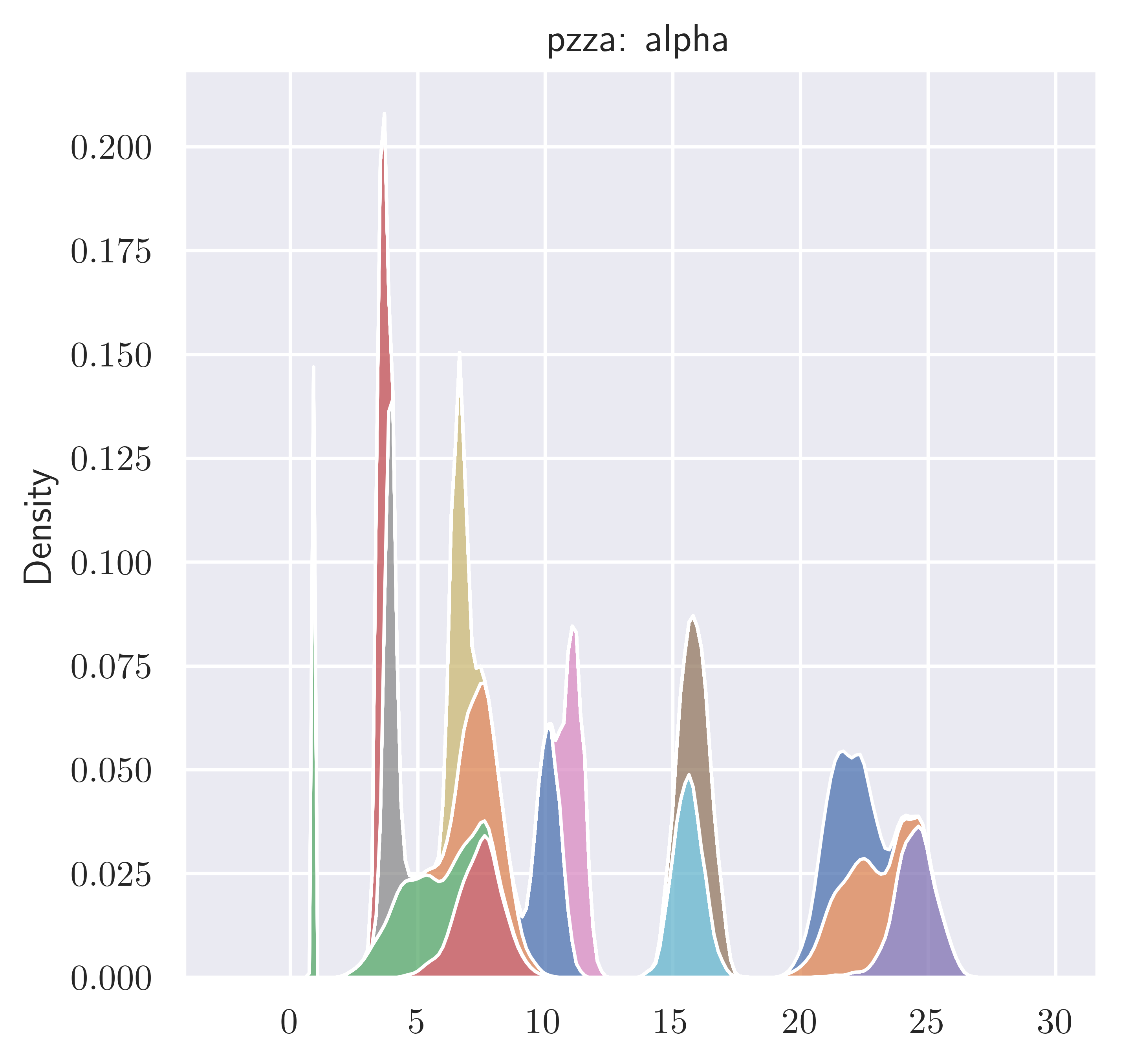}
	\label{fig:pzza_alpha}
}
\subfloat[PZZA $\beta$]{
	\includegraphics[scale=0.5]{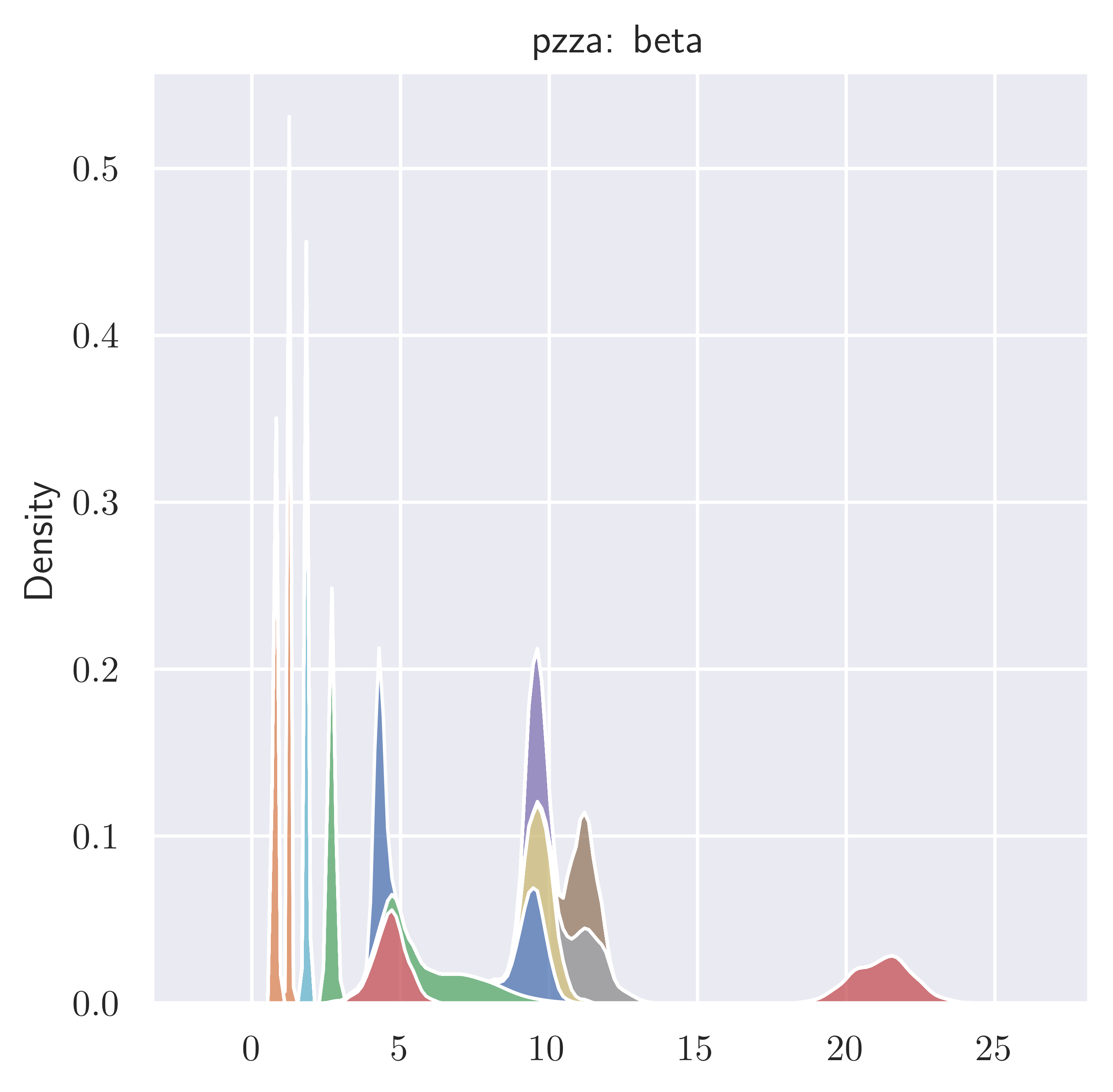}
	\label{fig:pzza_beta}
}
\caption{Inferred distributions over the mixture component distribution parameters $\alpha$ and $\beta$.}
\label{fig:alphabeta}
\end{figure}

\begin{figure}[th!]
\centering
\subfloat[AAPL $r$]{
	\includegraphics[scale=0.5]{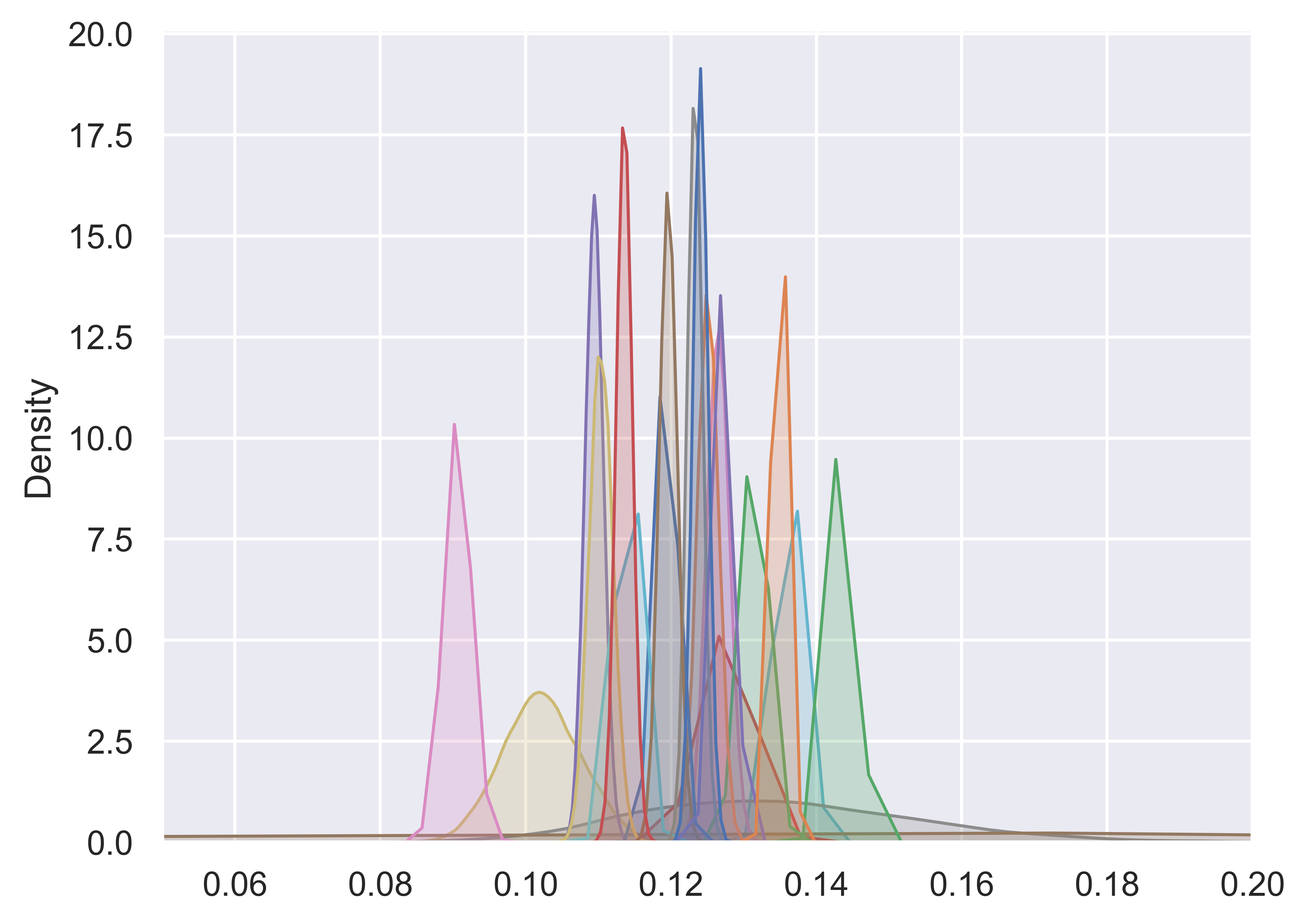}
	\label{fig:aapl_r}
}
\subfloat[AAPL $\tau$]{
	\includegraphics[scale=0.5]{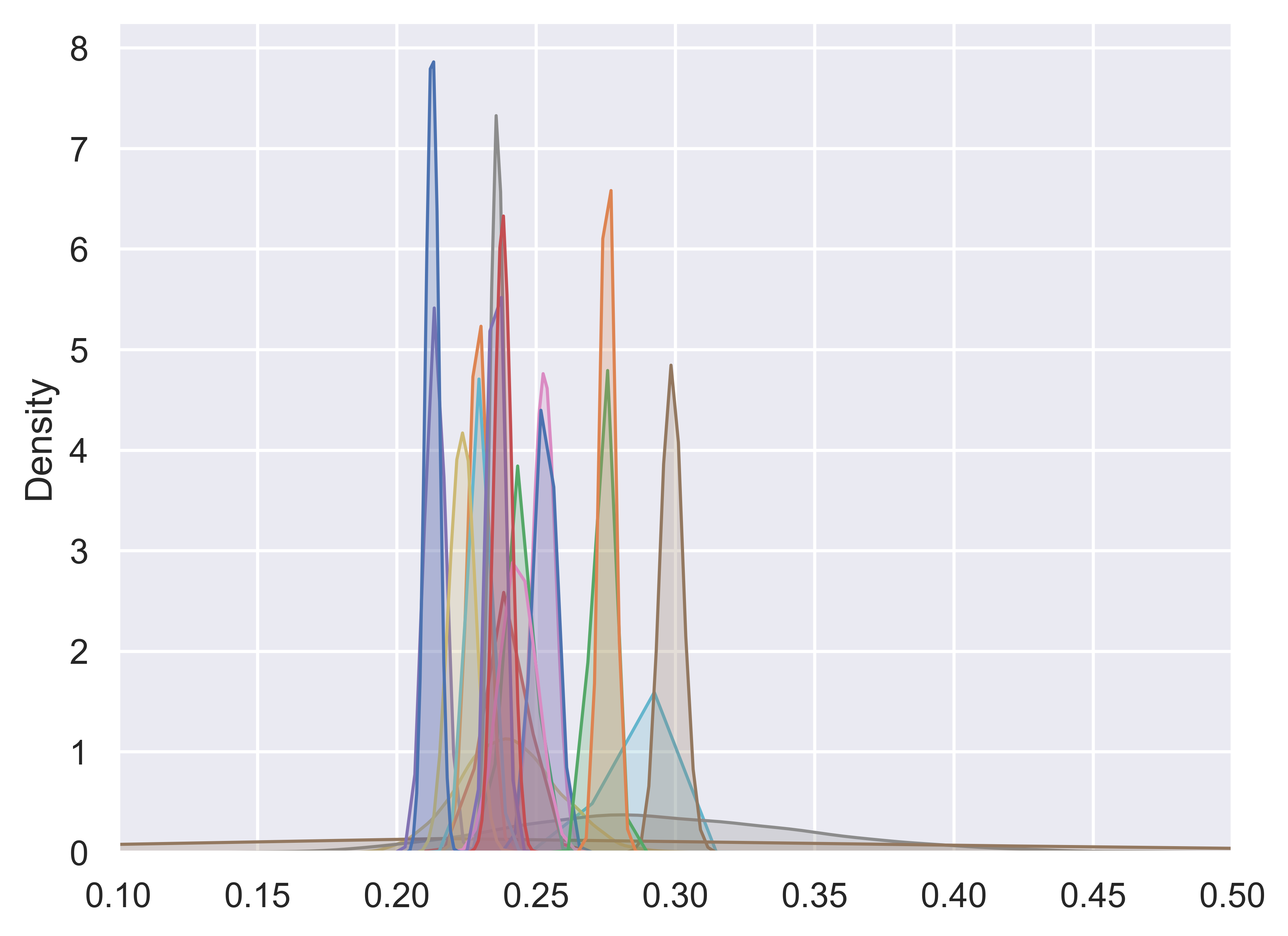}
	\label{fig:aapl_tau}
}\\
\subfloat[IRBT $r$]{
	\includegraphics[scale=0.5]{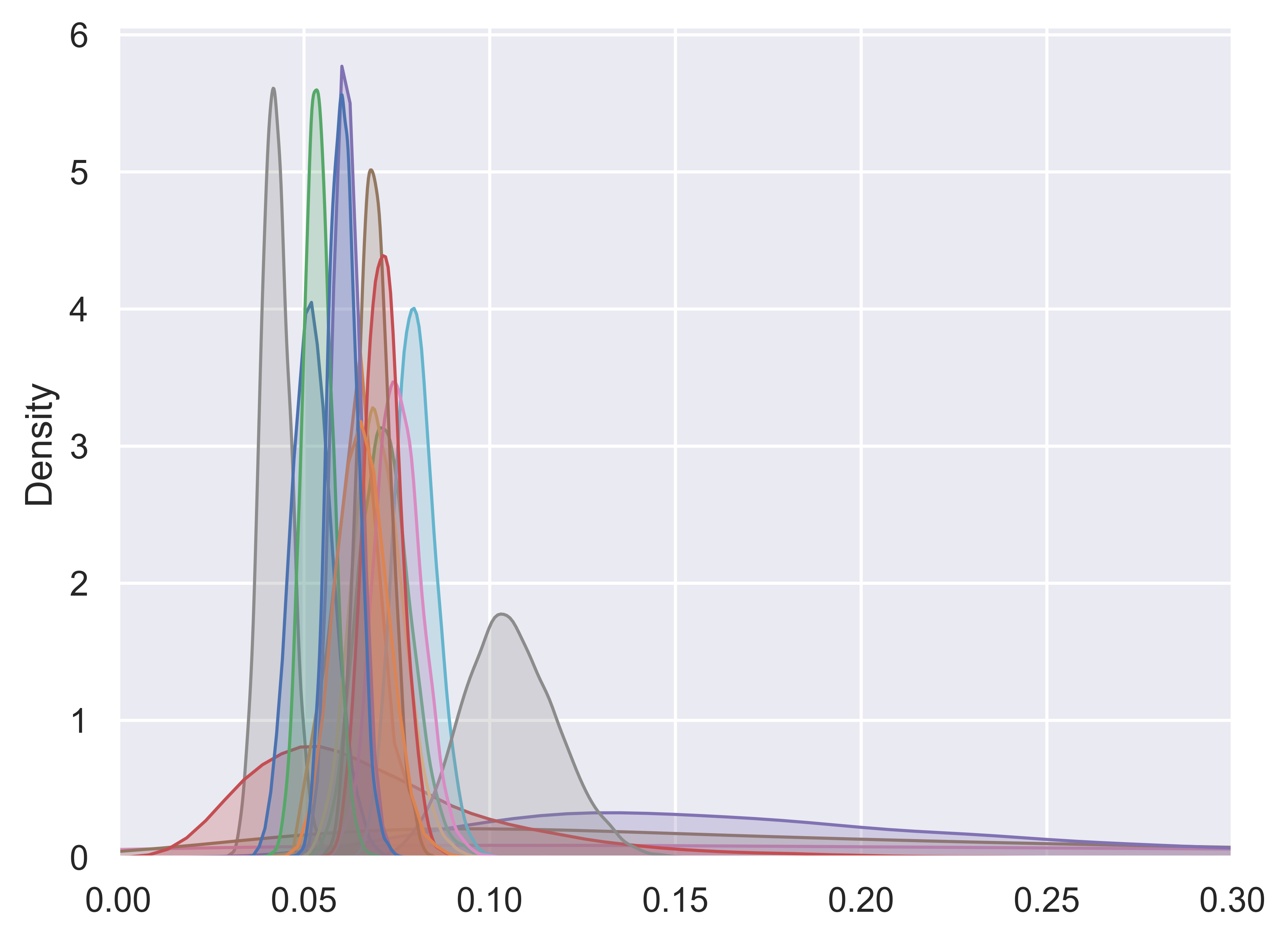}
	\label{fig:irbt_r}
}
\subfloat[IRBT $\tau$]{
	\includegraphics[scale=0.5]{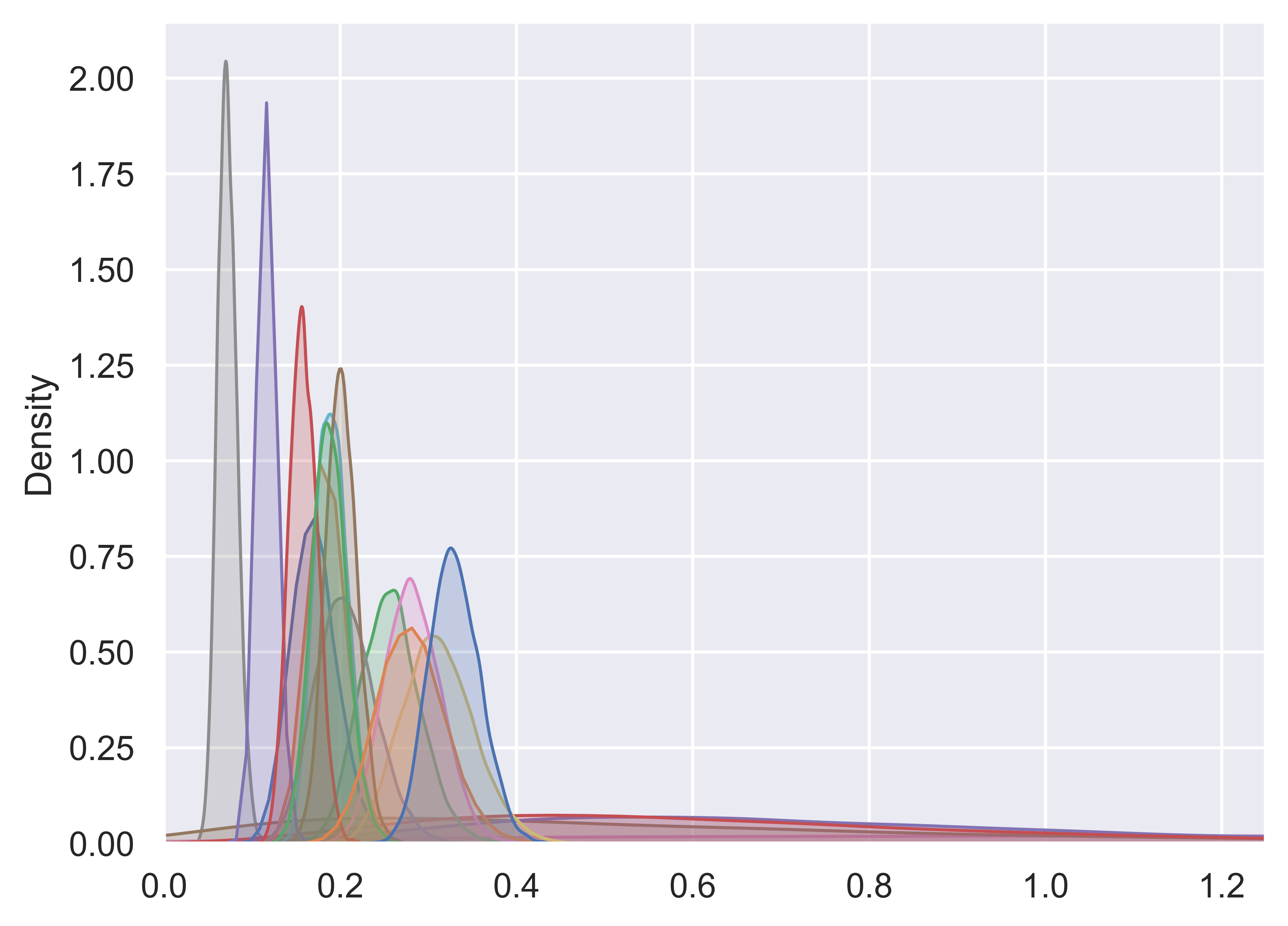}
	\label{fig:irbt_tau}
}\\
\subfloat[PZZA $r$]{
	\includegraphics[scale=0.5]{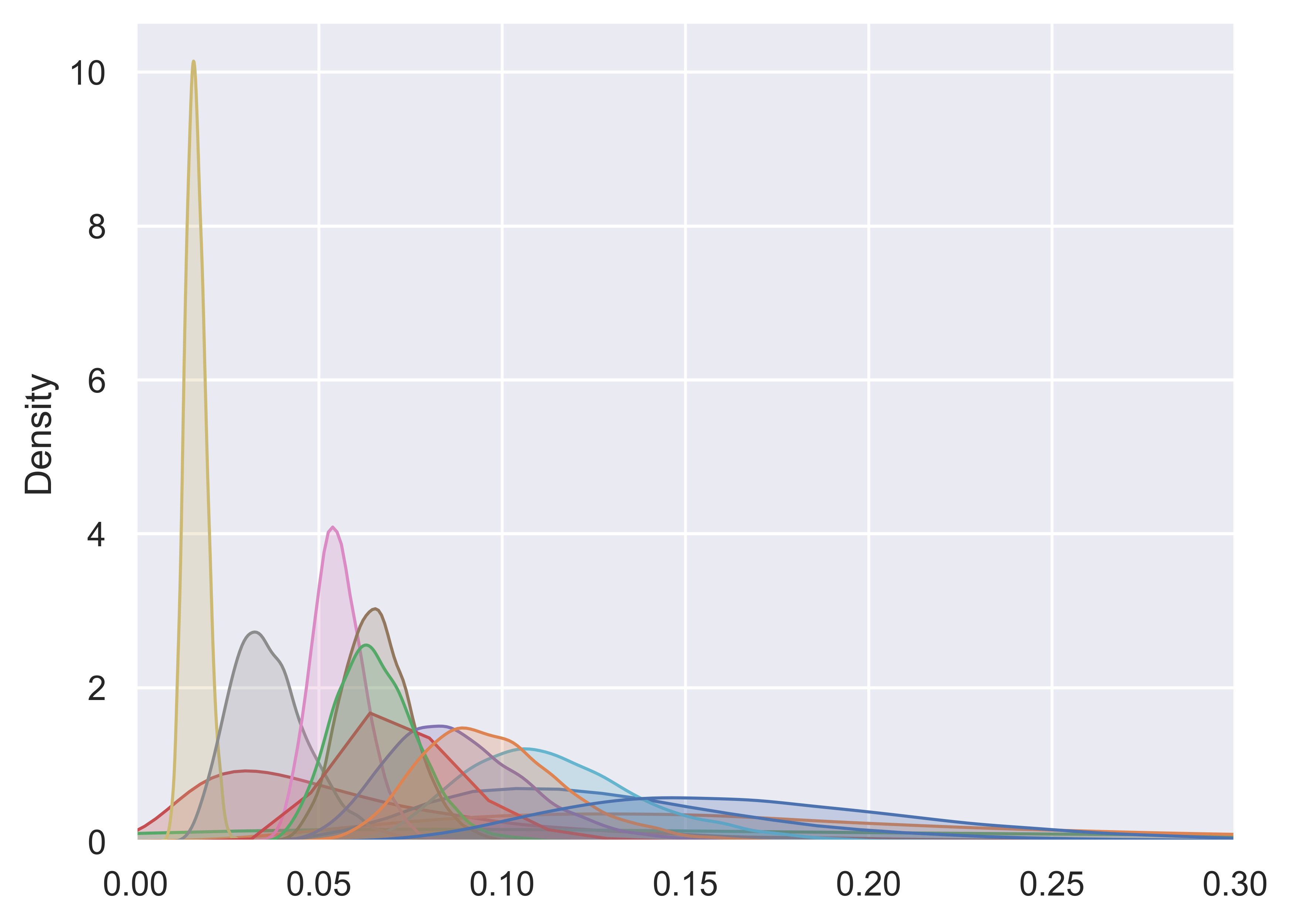}
	\label{fig:pzza_r}
}
\subfloat[PZZA $\tau$]{
	\includegraphics[scale=0.5]{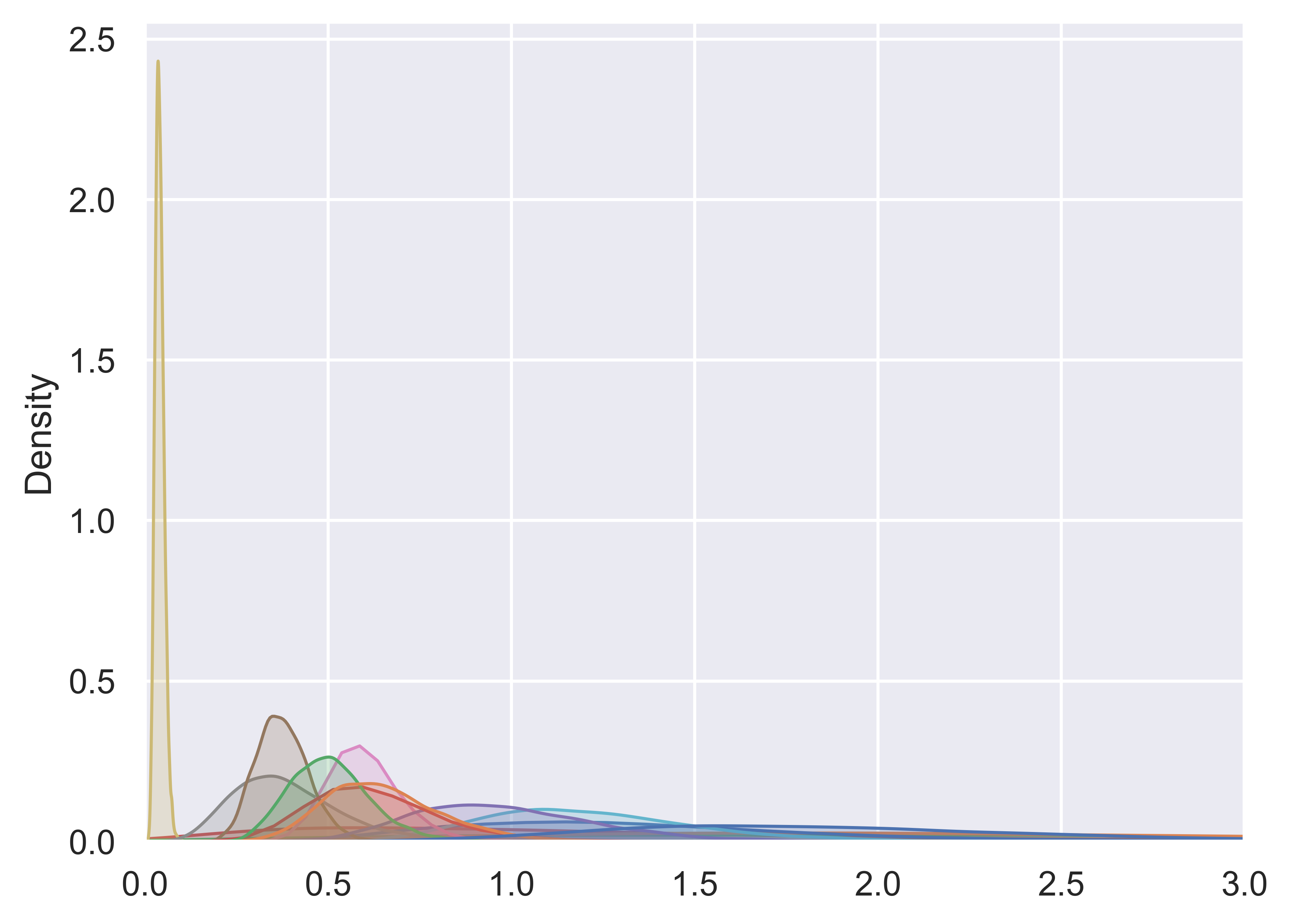}
	\label{fig:pzza_tau}
}
\caption{Inferred distributions over the parameters $r$ and $\tau$, which characterize the distribution over trade sizes.}
\label{fig:rtau}
\end{figure}

The inferred distributions over the parameters $\alpha$ and $\beta$ of the mixture component distributions are displayed in \cref{fig:alphabeta}, and the distributions over the parameters $r$ and $\tau$ characterizing the distribution over the trade sizes are displayed in \cref{fig:rtau}. These plots were generated from 10,000 MCMC samples (collected following the burn in) where all other parameters were held fixed at the MAP setting used to generate \cref{fig:results}.
Interestingly, for AAPL we see little overlap between the distributions of the parameters $\alpha$ and $\beta$ for each component; these distributions are very peaked as AAPL has by far the largest number of trades. 
The distributions of $\alpha$ and $\beta$ for IRBT and PZZA are broader and overlap across the components, with usually only a few distributions being very peaked. Similar statements appear to hold for the parameters $r$ and $\tau$.

\section{Conclusion}

The hierarchical Poisson process model is an appropriate choice for daily financial volume trading curves and learns a smooth inhomogenous Poisson process intensity function that shares components across different days. Inference via MCMC is fast for even the largest of stocks (like Apple) using the Dirichlet process slice sampling framework. The model may naturally extend to couple the mixture components across stocks and further up to broader groupings of the stocks, for example, to market segments. The slice sampling procedure readily extends correspondingly, though one may want to consider a more efficient approach than slice sampling for the mixture component parameters, such as Hamiltonian Monte Carlo, which would move these parameters and mix the Markov chain faster.
This hierarchical Poisson process model is novel and we hypothesize that it may be an appropriate model for a variety of other datasets that are naturally grouped into collections of event times.

\begin{funding}
AP and LFJ were supported by supported in part by the grants RGC-GRF 16300217 and T31-604/18-N of the HKSAR.
\end{funding}

\bibliographystyle{imsart-nameyear} 
\bibliography{hpp-paper}       

\end{document}

%% file: defs.tex
\def\[#1\]{\begin{align}#1\end{align}}

\newcommand{\dee}{\,\mathrm{d}}

\DeclareMathOperator{\gammadist}{gamma}
\DeclareMathOperator{\betadist}{beta}
\DeclareMathOperator{\Dirichlet}{Dirichlet}
\DeclareMathOperator{\Poisson}{Poisson}

\DeclareMathOperator{\nbdist}{neg-bin}
\DeclareMathOperator{\Uniform}{U}

\DeclareMathOperator{\DPLAW}{DP}

\DeclareMathOperator{\GEMLAW}{GEM}

\newcommand{\Ical}{\mathcal{I}}
\newcommand{\Acal}{\mathcal{A}}
\newcommand{\Bcal}{\mathcal{B}}

\renewcommand{\Pr}{\mathbb{P}}
\newcommand{\given}{\mid}
\newcommand{\dist}{\ \sim\ }

\newcommand{\Indicator}[1]{\mathbbm{1}{\{#1\}}}

\newcommand{\defas}{\vcentcolon=}  
\newcommand{\iid}{i.i.d.}
\newcommand{\as}{a.s.}
\newcommand{\pmf}{p.m.f.}

\newcommand{\PosReals}{\mathbb{R}_{>0}}
\newcommand{\PosInts}{\mathbb{Z}_{\ge 1}}
\newcommand{\conc}{\gamma}

%% file: hpp-paper.bbl
\begin{thebibliography}{16}

\bibitem[\protect\citeauthoryear{Adams, Murray and
  MacKay}{2009}]{adams2009tractable}
\begin{binproceedings}[author]
\bauthor{\bsnm{Adams},~\bfnm{R.~P.}\binits{R.~P.}},
  \bauthor{\bsnm{Murray},~\bfnm{I.}\binits{I.}} \AND
  \bauthor{\bsnm{MacKay},~\bfnm{D.~J.~C.}\binits{D.~J.~C.}}
(\byear{2009}).
\btitle{Tractable nonparametric {B}ayesian inference in {P}oisson processes
  with {G}aussian process intensities}.
In \bbooktitle{Proceedings of the 26th International Conference on Machine
  Learning}.
\end{binproceedings}
\endbibitem

\bibitem[\protect\citeauthoryear{Amini et~al.}{2019}]{amini2019exact}
\begin{barticle}[author]
\bauthor{\bsnm{Amini},~\bfnm{A.~A.}\binits{A.~A.}},
  \bauthor{\bsnm{Paez},~\bfnm{M.}\binits{M.}},
  \bauthor{\bsnm{Lin},~\bfnm{L.}\binits{L.}} \AND
  \bauthor{\bsnm{Razaee},~\bfnm{Z.~S.}\binits{Z.~S.}}
(\byear{2019}).
\btitle{Exact slice sampler for Hierarchical {D}irichlet Processes}.
\bjournal{arXiv preprint 1903.08829}.
\end{barticle}
\endbibitem

\bibitem[\protect\citeauthoryear{Ishwaran and James}{2004}]{ishwaran2004comp}
\begin{barticle}[author]
\bauthor{\bsnm{Ishwaran},~\bfnm{H.}\binits{H.}} \AND
  \bauthor{\bsnm{James},~\bfnm{L.~F.}\binits{L.~F.}}
(\byear{2004}).
\btitle{Computational Methods for Multiplicative Intensity Models Using
  Weighted Gamma Processes}.
\bjournal{Journal of the American Statistical Association}
\bvolume{99}
\bpages{175--190}.
\bdoi{10.1198/016214504000000179}
\end{barticle}
\endbibitem

\bibitem[\protect\citeauthoryear{Kalli, Griffin and
  Walker}{2011}]{kalli2011slice}
\begin{barticle}[author]
\bauthor{\bsnm{Kalli},~\bfnm{M.}\binits{M.}},
  \bauthor{\bsnm{Griffin},~\bfnm{J.~E.}\binits{J.~E.}} \AND
  \bauthor{\bsnm{Walker},~\bfnm{S.~G.}\binits{S.~G.}}
(\byear{2011}).
\btitle{Slice sampling mixture models}.
\bjournal{Statistics and Computing}
\bvolume{21}
\bpages{93--105}.
\end{barticle}
\endbibitem

\bibitem[\protect\citeauthoryear{Kingman}{1993}]{kingman1993}
\begin{bbook}[author]
\bauthor{\bsnm{Kingman},~\bfnm{J.~F.~C.}\binits{J.~F.~C.}}
(\byear{1993}).
\btitle{Poisson Processes}.
\bseries{Oxford Studies in Probability}.
\bpublisher{The Clarendon Press, Oxford University Press}.
\end{bbook}
\endbibitem

\bibitem[\protect\citeauthoryear{Kottas}{2006}]{kottas2006dirichlet}
\begin{binproceedings}[author]
\bauthor{\bsnm{Kottas},~\bfnm{A.}\binits{A.}}
(\byear{2006}).
\btitle{Dirichlet process mixtures of beta distributions, with applications to
  density and intensity estimation}.
In \bbooktitle{Workshop on Learning with Nonparametric Bayesian Methods, 23rd
  International Conference on Machine Learning (ICML)}.
\end{binproceedings}
\endbibitem

\bibitem[\protect\citeauthoryear{Kottas and
  Sans{\'o}}{2007}]{kottas2007bayesian}
\begin{barticle}[author]
\bauthor{\bsnm{Kottas},~\bfnm{A.}\binits{A.}} \AND
  \bauthor{\bsnm{Sans{\'o}},~\bfnm{B.}\binits{B.}}
(\byear{2007}).
\btitle{Bayesian mixture modeling for spatial {P}oisson process intensities,
  with applications to extreme value analysis}.
\bjournal{Journal of Statistical Planning and Inference}
\bvolume{137}
\bpages{3151--3163}.
\end{barticle}
\endbibitem

\bibitem[\protect\citeauthoryear{Lloyd et~al.}{2015}]{lloyd2015variational}
\begin{binproceedings}[author]
\bauthor{\bsnm{Lloyd},~\bfnm{C.}\binits{C.}},
  \bauthor{\bsnm{Gunter},~\bfnm{T.}\binits{T.}},
  \bauthor{\bsnm{Osborne},~\bfnm{M.}\binits{M.}} \AND
  \bauthor{\bsnm{Roberts},~\bfnm{S.}\binits{S.}}
(\byear{2015}).
\btitle{Variational inference for {G}aussian process modulated {P}oisson
  processes}.
In \bbooktitle{Proceedings of the 32nd International Conference on Machine
  Learning}
\bpages{1814--1822}.
\end{binproceedings}
\endbibitem

\bibitem[\protect\citeauthoryear{M{\o}ller, Syversveen and
  Waagepetersen}{1998}]{moller1998log}
\begin{barticle}[author]
\bauthor{\bsnm{M{\o}ller},~\bfnm{J.}\binits{J.}},
  \bauthor{\bsnm{Syversveen},~\bfnm{A.~R.}\binits{A.~R.}} \AND
  \bauthor{\bsnm{Waagepetersen},~\bfnm{R.~P.}\binits{R.~P.}}
(\byear{1998}).
\btitle{Log {G}aussian {C}ox processes}.
\bjournal{Scandinavian Journal of Statistics}
\bvolume{25}
\bpages{451--482}.
\end{barticle}
\endbibitem

\bibitem[\protect\citeauthoryear{Neal}{2003}]{neal2003slice}
\begin{barticle}[author]
\bauthor{\bsnm{Neal},~\bfnm{R.~M.}\binits{R.~M.}}
(\byear{2003}).
\btitle{Slice sampling}.
\bjournal{Annals of statistics}.
\end{barticle}
\endbibitem

\bibitem[\protect\citeauthoryear{Pitman}{2006}]{pitman2006combinatorial}
\begin{bbook}[author]
\bauthor{\bsnm{Pitman},~\bfnm{J.}\binits{J.}}
(\byear{2006}).
\btitle{Combinatorial Stochastic Processes: {E}cole d'{E}t{\'e} de
  {P}robabilit{\'e}s de {S}aint--{F}lour XXXII-2002}.
\bseries{Lecture Notes in Mathematics}.
\bpublisher{Springer}.
\end{bbook}
\endbibitem

\bibitem[\protect\citeauthoryear{Robert and Casella}{2013}]{robert2013monte}
\begin{bbook}[author]
\bauthor{\bsnm{Robert},~\bfnm{C.}\binits{C.}} \AND
  \bauthor{\bsnm{Casella},~\bfnm{G.}\binits{G.}}
(\byear{2013}).
\btitle{Monte {C}arlo statistical methods}.
\bpublisher{Springer Science \& Business Media}.
\end{bbook}
\endbibitem

\bibitem[\protect\citeauthoryear{Sethuraman}{1994}]{sethuraman1994constructive}
\begin{barticle}[author]
\bauthor{\bsnm{Sethuraman},~\bfnm{J.}\binits{J.}}
(\byear{1994}).
\btitle{A constructive definition of {D}irichlet priors}.
\bjournal{Statistica Sinica}
\bvolume{4}
\bpages{639--650}.
\end{barticle}
\endbibitem

\bibitem[\protect\citeauthoryear{Taddy and Kottas}{2012}]{taddy2012mixture}
\begin{barticle}[author]
\bauthor{\bsnm{Taddy},~\bfnm{M.~A.}\binits{M.~A.}} \AND
  \bauthor{\bsnm{Kottas},~\bfnm{A.}\binits{A.}}
(\byear{2012}).
\btitle{Mixture modeling for marked {P}oisson processes}.
\bjournal{Bayesian Analysis}
\bvolume{7}
\bpages{335--362}.
\end{barticle}
\endbibitem

\bibitem[\protect\citeauthoryear{Teh et~al.}{2006}]{tehhdp}
\begin{barticle}[author]
\bauthor{\bsnm{Teh},~\bfnm{Y.~W.}\binits{Y.~W.}},
  \bauthor{\bsnm{Jordan},~\bfnm{M.~I.}\binits{M.~I.}},
  \bauthor{\bsnm{Beal},~\bfnm{M.~J.}\binits{M.~J.}} \AND
  \bauthor{\bsnm{Blei},~\bfnm{D.~M.}\binits{D.~M.}}
(\byear{2006}).
\btitle{Hierarchical {D}irichlet Processes}.
\bjournal{Journal of the American Statistical Association}
\bvolume{101}
\bpages{1566--1581}.
\end{barticle}
\endbibitem

\bibitem[\protect\citeauthoryear{Walker}{2007}]{walker2007sampling}
\begin{barticle}[author]
\bauthor{\bsnm{Walker},~\bfnm{S.~G.}\binits{S.~G.}}
(\byear{2007}).
\btitle{Sampling the {D}irichlet mixture model with slices}.
\bjournal{Communications in Statistics -- Simulation and Computation}
\bvolume{36}
\bpages{45--54}.
\end{barticle}
\endbibitem

\end{thebibliography}
